\documentclass[aps,twocolumn,floats,prd,nofootinbib,superscriptaddress,10pt,longbibliography]{revtex4-1}
\usepackage[utf8]{inputenc}
\usepackage{graphicx}
\usepackage{dcolumn}
\usepackage{bm}
\usepackage{hyperref}
\usepackage{lipsum}
\usepackage{xcolor}
\usepackage{calc}
\usepackage{comment}
\usepackage{float}
\usepackage{diagbox}
\usepackage{soul}
\usepackage{ulem}
\usepackage{amsmath}
\usepackage{multirow}

\makeatletter
\newcommand\footnoteref[1]{\protected@xdef\@thefnmark{\ref{#1}}\@footnotemark}
\makeatother

\begin{document}

\preprint{APS/123-QED}

\title{Towards alleviating the $H_0$ and $S_8$ tensions with Early Dark Energy -- Dark Matter drag}
\author{Th\'eo Simon}
\affiliation{Laboratoire Univers \& Particules de Montpellier (LUPM), CNRS \& Universit\'e de Montpellier (UMR-5299),Place Eug\`ene Bataillon, F-34095 Montpellier Cedex 05, France}
\author{Tal Adi}
\affiliation{Physics Department, Ben-Gurion University of the Negev, Beersheba, Israel}
\author{Jos\'e Luis Bernal}
\affiliation{Instituto de Física de Cantabria (IFCA), CSIC-Univ. de Cantabria, Avda. de los Castros s/n, E-39005 Santander, Spain}
\author{Ely D. Kovetz}
\affiliation{Physics Department, Ben-Gurion University of the Negev, Beersheba, Israel}
\author{Vivian Poulin}
\affiliation{Laboratoire Univers \& Particules de Montpellier (LUPM), CNRS \& Universit\'e de Montpellier (UMR-5299),Place Eug\`ene Bataillon, F-34095 Montpellier Cedex 05, France}
\author{Tristan L.~Smith}
\affiliation{Department of Physics and Astronomy, Swarthmore College, Swarthmore, PA 19081, USA}

\begin{abstract}
Early dark energy, an additional component of dark energy active in the decade of redshift before recombination, has emerged as one of the most effective models at reducing the ``$H_0$ tension'' between direct measurement of the Hubble parameter $H_0$ in the late-universe and the $\Lambda$CDM prediction when calibrated on {\it Planck}. However, it requires a slight increase in the dark matter density $\omega_{\rm cdm}$ and primordial tilt $n_s$ that worsens the ``$S_8$ tension'' between measurements of weak gravitational lensing at low redshifts and the {\it Planck}/$\Lambda$CDM prediction. Using a phenomenological fluid model, we investigate whether the inclusion of a drag term between dark matter and early dark energy can compensate for the effect of the increase in power at small-scales, such that both $H_0$ and $S_8$ tensions are simultaneously alleviated. We find that this works if the drag term is dynamically relevant in the post-recombination universe. However, a drag term active before or just around the time at which the early dark energy contribution to the energy density is maximum is  significantly constrained due to its impact on the matter perturbations before recombination, and the subsequent modifications to the cosmic microwave background power spectra.

\end{abstract}

\maketitle
\section{Introduction}

The increasing precision of observations has revealed discrepancies within the concordance $\Lambda$-cold-dark-matter ($\Lambda$CDM) cosmological model.
Indeed, measurements of some parameters (in particular the $H_0$ and $S_8$ parameters) take different values for different observables and experiments (see {\it e.g.}, Ref.~\cite{Abdalla:2022yfr} for a recent review). Though the possibility of unknown systematic effects explaining these discrepancies is extensively studied (see {\it e.g.}, \cite{Freedman:2021ahq,Riess:2021jrx,Amon:2022ycy,Arico:2023ocu} for discussion), it is possible that these ``cosmic tensions'' indicate departure from the consensus $\Lambda$CDM paradigm, with new dynamical effects sourcing these apparent discrepancies.

On the one hand, early dark energy (EDE), an additional component of dark energy (DE) active in the decade of redshift before recombination \cite{Karwal:2016vyq,Poulin:2018cxd,Smith:2019ihp,Poulin:2023lkg}, has emerged as a promising model to explain the ``Hubble tension'' between direct measurement of the Hubble parameter in the late-universe and the $\Lambda$CDM prediction when calibrated on {\it Planck} (see Refs.~\cite{DiValentino:2021izs,Schoneberg:2021qvd,Poulin:2023lkg} for reviews of the models which have been proposed to address the Hubble tension). However, as cosmological observations strongly constrain the expansion history at late times (effectively fixing $\Omega_m$), the increase in $H_0$ necessarily leads to a large increase in $\omega_{\rm m}\equiv \Omega_m h^2$, where $h=100H_0$km/s/Mpc.
Therefore, the resulting inferred value of $S_8$ increases as well \cite{Jedamzik:2020krr,Poulin:2024ken}. While EDE can compensate the effect of $\omega_{\rm cdm}$ in the cosmic microwave background (CMB), it has been found that typically in cosmologies where the Hubble tension is resolved (such as with EDE), the $S_8$ tension can be exacerbated \cite{Hill:2020osr,Murgia:2020ryi}.   

On the other hand, the $S_8$ tension is an indication of a suppression of power at small-scales, which could be independently resolved thanks to some new dark matter (DM) properties, such as decay or interaction with baryons, dark radiation or DE (see \textit{e.g.}, \cite{Lesgourgues:2015wza,Buen-Abad:2015ova,Chacko:2016kgg,Buen-Abad:2017gxg,Heimersheim:2020aoc,DiValentino:2019ffd,Lucca:2021dxo,Abellan:2020pmw,DiValentino:2020vvd,Bansal:2021dfh,Baldi:2016zom,Kumar:2017bpv,Asghari:2019qld,BeltranJimenez:2020qdu,Figueruelo:2021elm,BeltranJimenez:2021wbq,Poulin:2022sgp}), or some non-linear effect \cite{Amon:2022azi,Stahl:2024stz}. In particular, Refs.~\cite{Pourtsidou:2016ico,Baldi:2016zom,Kumar:2017bpv,Asghari:2019qld,BeltranJimenez:2020qdu,Figueruelo:2021elm,BeltranJimenez:2021wbq,Poulin:2022sgp} suggests that a drag between DM and DE can slow down the falling of DM into gravitational potential wells and thereby suppresses the growth of power. 

In this paper, we explore whether a similar effect could have occurred in the pre-recombination universe, between an EDE component and dark matter, and investigate if such a model can simultaneously explain the $H_0$ and $S_8$ tensions. Other attempts in that direction have been explored in the literature \cite{McDonough:2021pdg,Karwal:2021vpk,Liu:2023haw,Liu:2023rvo,Liu:2023wew,Garcia-Arroyo:2024tqq}, where a coupling between scalar-field EDE and dark matter is introduced at the Lagrangian level (see \textit{e.g.}, \cite{Pourtsidou:2013nha} for a general discussion). However, the difficulty in such approaches is that there is no unique choice of coupling, and the phenomenology may be very different depending on the choice of coupling and scalar field potential. For instance, Refs.~\cite{McDonough:2021pdg,Karwal:2021vpk,Liu:2023rvo} conclude that a Yukawa-like coupling would mostly increase $S_8$, as it leads to a stronger gravitational constant at early times (see also Ref.~\cite{Bean:2008ac}).

Rather than focusing on specific models, we explore a phenomenological approach where the drag is introduced at the level of the linearly perturbed Euler equation, enforcing momentum conservation between an EDE fluid and the DM fluid. As such, our goal is to extract general properties that models must fulfill in order to alleviate both tensions, and provide guidelines for model builders. The caveat is that some aspects of the dynamics included in any specific model may be missing from our phenomenological approach. Given the tight constraints on the properties of dark matter between recombination and today \cite{Ilic:2020onu}, we expect that any additional effects in specific models beyond modifications to the linearly perturbed Euler equation will lead to further constraints implying that this is likely a conservative approach. 

Our paper is structured as follows. In Sec.~\ref{sec:model}, we detail our modeling of the DM-EDE drag and describe the impact on the CMB and matter power spectra. In Sec.~\ref{sec:analysis}, we perform a series of Bayesian analyses in light of up-to-date compilation of cosmological data to assess whether the model can alleviate both tensions, and we conclude in Sec.~\ref{sec:conclusions}.
In addition, in App.~\ref{sec:discussion}, we discuss a potential theoretical embedding of our phenomenological model and the issues it currently faces, while in App.~\ref{app:ede_fluid} we review the EDE model we consider in this work. Finally, in Apps.~\ref{app:early_and_local} and~\ref{app:chi2}, we provide additional material from our analyses.

\section{A drag between DM and EDE}
\label{sec:model}
\subsection{Pure momentum transfer model}

There are several possible ways to include a drag between DM and DE. A common phenomenological choice is to model it as an exchange of momentum that modifies the Euler equations with a term $\propto (\theta_c-\theta_\phi)$, where $\theta_X$ is the velocity perturbation in species $X$. However, as we now discuss, this simple phenomenological choice does not follow from more detailed models. 

If the EDE is modeled as a scalar field, a well-defined way is to introduce a coupling in the scalar field Lagrangian, following \textit{e.g.} Refs.~\cite{Pourtsidou:2016ico,Pourtsidou:2013nha}. As discussed in the literature \cite{McDonough:2021pdg,Karwal:2021vpk,Liu:2023haw,Liu:2023rvo,Liu:2023wew,Garcia-Arroyo:2024tqq}, the choice of coupling can generally lead to a broad phenomenology. 
The simplest coupling that leads to a pure momentum transfer model can be described by the Lagrangian (in the formalism of Ref.~\cite{Pourtsidou:2013nha}):
\begin{equation}
    \mathcal{L}=\frac{1}{2}\nabla_\mu\phi\nabla^\mu\phi+\beta \left(u^\mu\nabla_\mu\phi\right)^2 - V(\phi), \label{eq:lagrangian}
\end{equation}
where $\phi$ is a quintessence scalar field, $u^\mu$ is the four-velocity of the fluid, $V(\phi)$ is the potential and $\beta$ is a coupling constant.
Such a coupling leaves the DM continuity equation unaffected, but leads to a modified Euler equation for the CDM: 
\begin{equation}
    \dot{\theta}_{\rm DM}+\frac{\dot{a}}{a}\theta_{\rm DM}=-2\beta\frac{\left(\frac{\ddot{\phi}_{0}}{a}+2\frac{\dot{a}}{a}\frac{\dot{\phi}_{0}}{a}\right)\phi_{1}+\frac{\dot{\phi}_{0}}{a}\dot{\phi}_{1}}{a\left(\overline{\rho}_{\rm DM}-2\beta\frac{\dot{\phi}_{0}^{2}}{a^{2}}\right)}k^{2} \, ,
\end{equation}
where the scalar field $\phi$ is expressed in terms of a homogeneous contribution $\phi_0$ and a linear perturbation $\phi_1$. Let us note that we use \textit{overdot} to denote conformal time derivative. 
The coupling of Eq.~\eqref{eq:lagrangian} also leads to modified Klein Gordon equations for the scalar field, which are explicitly given in App.~\ref{sec:discussion}. 
It is instructive to write the Euler equation for both species in terms of the field fluid quantities (expressed in App.~\ref{sec:discussion}), and in the limit of weak momentum coupling (\textit{i.e.}, when $\left|\beta\right|\ll 1$). In that case, we find (using the notation of Ref.~\cite{Ma:1995ey})
\begin{align}\label{eq:EulerSimp}
    \dot{\theta}_{\rm DM}\approx&-\frac{\dot{a}}{a}\theta_{c}-2\beta\left(\frac{\delta\rho_{\phi}}{\overline{\rho}_{c}}k^{2}+3\frac{\dot{a}}{a}\frac{\overline{\rho}_{\phi}+\overline{P}_{\phi}}{\overline{\rho}_{c}}\left(1-c_{\phi}^{2}\right)\theta_{\phi}\right) \, ,\nonumber\\
    \dot{\theta}_{\phi}\approx& 2\frac{\dot{a}}{a}\theta_{\phi}+\frac{\delta\rho_{\phi}}{\overline{\rho}_{\phi}+\overline{P}_{\phi}}k^{2}+6\beta\frac{\dot{a}}{a}\theta_{c} \, ,
\end{align}
where $\overline{\rho}_{\phi}$ and $\overline{P}_{\phi}$ are respectively the energy and pressure of the scalar field, and where $\delta\rho_{\phi}$ and $c_\phi^2\equiv \dot{\overline{P}}_\phi/\dot{\overline{\rho}}_\phi$  are respectively the scalar field density perturbation and the adiabatic sound speed.
One can see that such a theory derived from first principles does not lead to a drag term $\propto (\theta_c-\theta_\phi)$, unless $\delta \rho_\phi$ is negligible, which is only achieved close to the slow-roll limit, when the EDE scalar field behaves like a cosmological constant. This has been studied in Ref.~\cite{Pourtsidou:2016ico} in the context of DM-DE scattering. 

However, an EDE field does not always behave like a cosmological constant. In order to see the impact of a minimal change, here we limit the impact of the coupling to a pure drag term between DM and EDE through a phenomenological approach in which we model EDE and DM as fluids. We introduce the drag term directly at the level of the Euler equations, enforcing momentum conservation between the two fluids. 
Further discussion about this detailed model and how its phenomenology may differ from a pure drag term is presented in App.~\ref{sec:discussion}.

\subsection{A phenomenological model of DM$-$EDE drag}

To include the drag in the fluid formalism, working in Newtonian gauge, one simply needs to modify the evolution equations for the velocity divergences $\theta$ as
\begin{eqnarray}\label{eq:EulerPheno}
    \dot{\theta}_{\rm DM} &  =  & -\frac{\dot{a}}{a}\theta_{\rm DM}+k^2\psi +\Gamma_{\rm DM/EDE}(a)(\theta_{\rm EDE}-\theta_{\rm DM}) \, ,\nonumber\\
    \dot{\theta}_{\rm EDE}  &  =  &   -(1-3c_{s,{\rm EDE}}^2)\frac{\dot{a}}{a}\theta_{\rm EDE} +\frac{k^2c_{s,{\rm EDE}}^2}{(1+w_{\rm EDE})}\delta_{\rm EDE}\nonumber\\
   & &
    +k^2\psi - \Gamma_{\rm DM/EDE}(a)R(\theta_{\rm EDE}-\theta_{\rm DM}) \, ,
\end{eqnarray}
where $a$ is the scale factor, $\psi$ is the gravitational potential, $c_{s,{\rm DE}}\equiv1$ is the EDE sound speed, and $w_{\rm EDE}$ is the EDE equation-of-state parameter. The coefficient 
\begin{equation}
    R=\frac{\bar\rho_{\rm DM}(a)}{(1+w_{\rm EDE})\bar\rho_{\rm EDE}(a)}\,,
\end{equation}
where $\bar{\rho}_i$ are the mean proper energy densities of DM and EDE, ensures momentum conservation. Our choice of EDE model, that was introduced elsewhere and is based on the modified axion-like potential~\cite{Poulin:2018dzj,Poulin:2018cxd,Smith:2019ihp} is reviewed in App.~\ref{app:ede_fluid}. It is chosen because it has been found to be among the most successful EDE models to reduce the Hubble tension \cite{Poulin:2023lkg} and is favored by ACT CMB data \cite{Hill:2021yec,Poulin:2021bjr}, although the latest {\it Planck} data provide tighter constraints \cite{Efstathiou:2023fbn}. We do not anticipate major changes in our conclusions with a different fluid model, see \textit{e.g.} Ref.~\cite{Lin:2019qug,Niedermann:2019olb}.

The question is then how to parameterize the EDE-DM interaction rate $\Gamma_{\rm DM/EDE}(a)$. 
We explore three different choices, motivated by testing different functional forms tied to whether the interaction is active {\it before}, {\it around} or {\it after} the moment where EDE component makes its maximum contribution to the total energy density. 
First, we make use of the parameterization
\begin{equation}
    \Gamma^{\rm local}_{\rm DM/EDE}(a)=\beta f_{\rm EDE}(a) \,,
\end{equation}
which corresponds to an interaction rate that is relevant only when the EDE contribution to the overall energy density is relevant. We will dub this parameterization the ``localized'' interaction rate, to emphasize that its contribution is localized around the time at which EDE reaches its peak.
Second, we study the case of late-time interaction between DM and EDE, where the interaction rate is relevant only when (late) DE starts dominating the energy budget\cite{Simpson:2010vh,Asghari:2019qld,Poulin:2022sgp}:
\begin{equation}
    \Gamma^{\rm late}_{\rm DM/EDE}(a)=\frac{a \beta }{\bar{\rho}_{\rm DM}(a)}\rho_{\rm EDE,0} \,,
\end{equation}
where $\rho_{\rm EDE,0}$ (the EDE energy density today) appears explicitly to ensure that $\Gamma^{\rm late}_{\rm DM/EDE}(a)\to 0$ in the absence of EDE. Note that the rate is $\propto a^4$, such that the interaction is {\it not} active when the EDE component makes its largest contribution. Rather, it becomes relevant at much later time, around the time at which DE becomes relevant \cite{Poulin:2022sgp}. 
Such a drag may arise naturally from a model in which EDE transitions to late-time dark energy; we leave the exploration of this possibility for future work.
At this stage, we simply test whether the early-time energy injection from the EDE component, together with a late-time drag, can help resolving both tensions.
Finally, it is also possible that the drag was active in the early universe, and switches off when the EDE contribution decays away. 
This resembles for instance the scenario of Ref.~\cite{Joseph:2022jsf}, where an initially relativistic dark radiation (DR) interacts with dark matter, until it decouples from the cosmic bath implying the cessation of the interaction. 
We consider a similar scenario, where 
\begin{equation}
    \Gamma^{\rm early}_{\rm DM/EDE}(a)=a^{-5}\beta f_{\rm EDE}(a) \,.
\end{equation}
Note that the scaling with $a$ is arbitrary in this phenomenological approach, and we set it such that the ratio $\Gamma
/(aH)$ is constant during radiation domination, similarly to what happens in the DM-DR scenario of Ref.~\cite{Joseph:2022jsf}. 
We checked that different scaling before or after the transition does not affect our conclusions.

\begin{figure}
    \centering
    \includegraphics[width=\columnwidth]{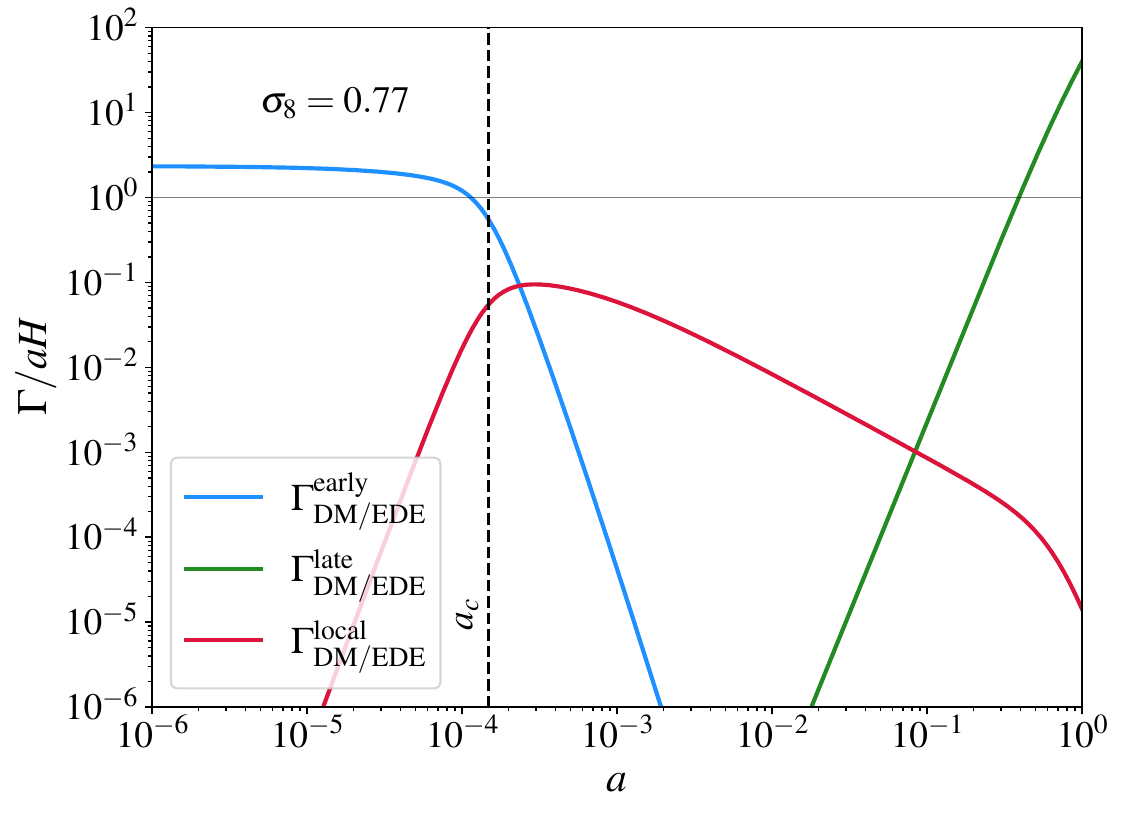}
    \caption{Scaling of the interaction rate with the scale factor $a$ for the three different parameterizations considered in this work. All three models use the same EDE parameters, and the interaction rates are normalized such that $\sigma_8 \simeq 0.77$.}
    \label{fig:Gamma}
\end{figure}

In the end, for each of the three parameterizations, the only additional free parameter of the model compared to standard EDE is $\beta$, the overall amplitude of the interaction.
We show the redshift dependence of each interaction rate in Fig.~\ref{fig:Gamma}, using an EDE model given by $f_{\rm EDE}(z_c)=0.1$, $\log_{10}(z_c)=-3.8$, $\Theta_i=2.9$, and values of $\beta$ that are adjusted such that $\sigma_8\simeq 0.77$. 
These correspond approximately to the EDE best-fit parameters extracted from an analysis that combines Planck+BAO+SN1a+S$H_0$ES data (that we will call ``${\cal DH}$'' in next section), while the values of $\beta$ are adjusted such that each model approximately give value of $S_8$ equal to that measured by weak lensing experiments.
One can see that, as expected, the early model shows a large  interaction rate (namely, $\Gamma > aH$) at $z>z_c$, and quickly drop afterwards. 
The interaction rate of the local model peaks around $z_c$, but is quickly negligible before and after $z_c$. 
Note that in this case, given the condition of $\sigma_8\simeq 0.77$, $\Gamma < aH$ at all times. 
Finally, the interaction rate of the late model increases with $a$ and is only relevant at much later times, closer to dark-energy domination, after which $\Gamma \gg aH$.

\begin{figure}[h!]
    \centering
     \includegraphics[width=0.8
\columnwidth]{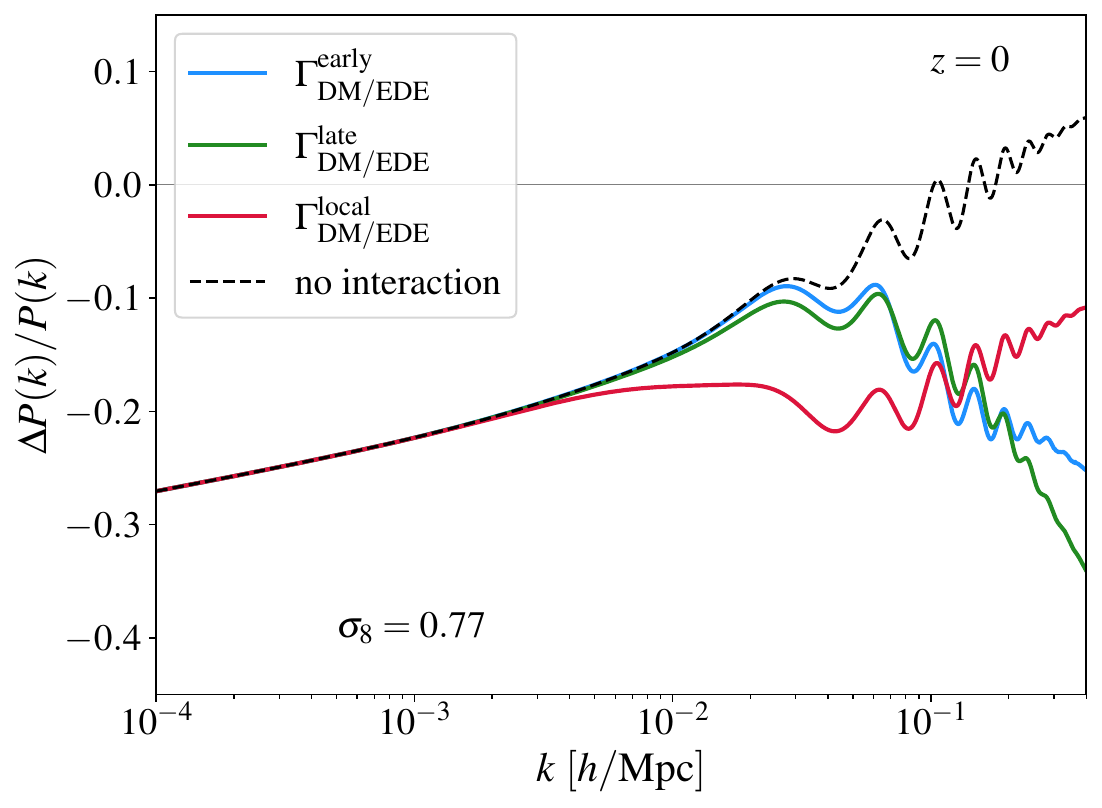}
    \includegraphics[width=0.8\columnwidth]{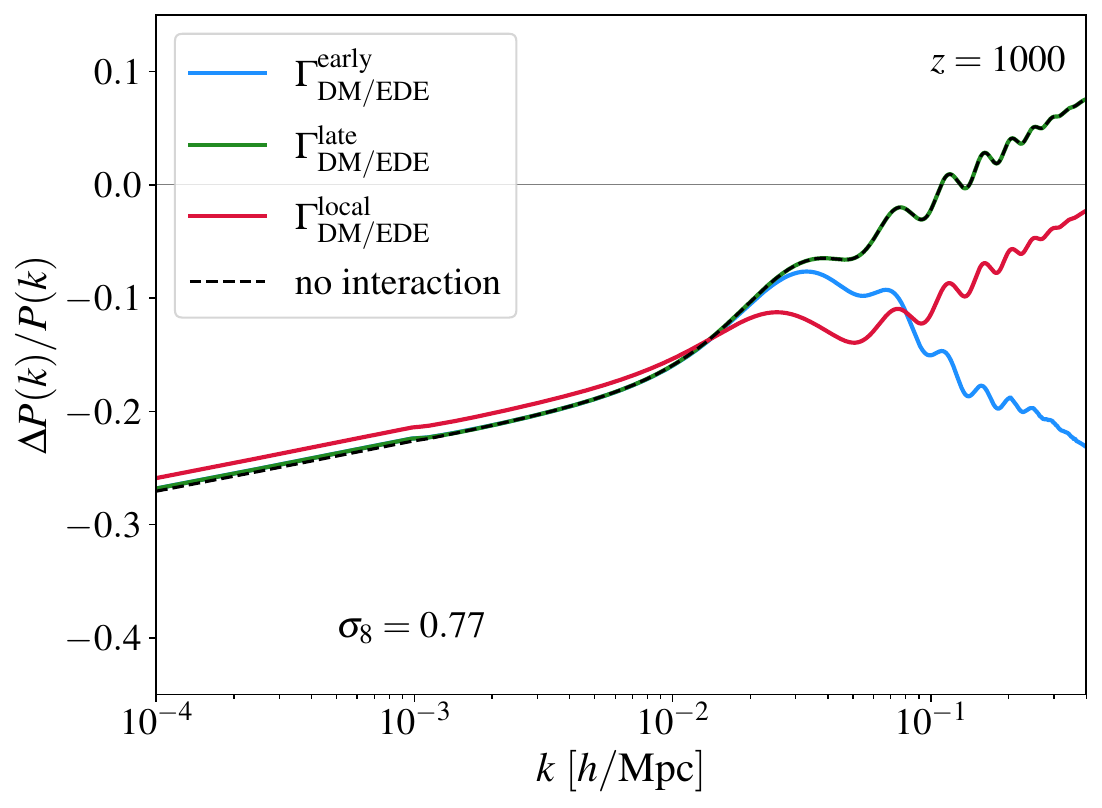}

\caption{Residuals of the matter power spectrum between EDE and $\Lambda$CDM, with and without the drag term. We illustrate the effect of the three parametrizations of $\Gamma$, normalized such that $\sigma_8 \simeq 0.77$. }
    \label{fig:pk}
\end{figure}

\subsection{Impact of EDE-DM drag on the matter and CMB power spectra} \label{sec:impact_CMB_LSS}

\begin{figure}[h!]
    \centering
    \includegraphics[width=\columnwidth]{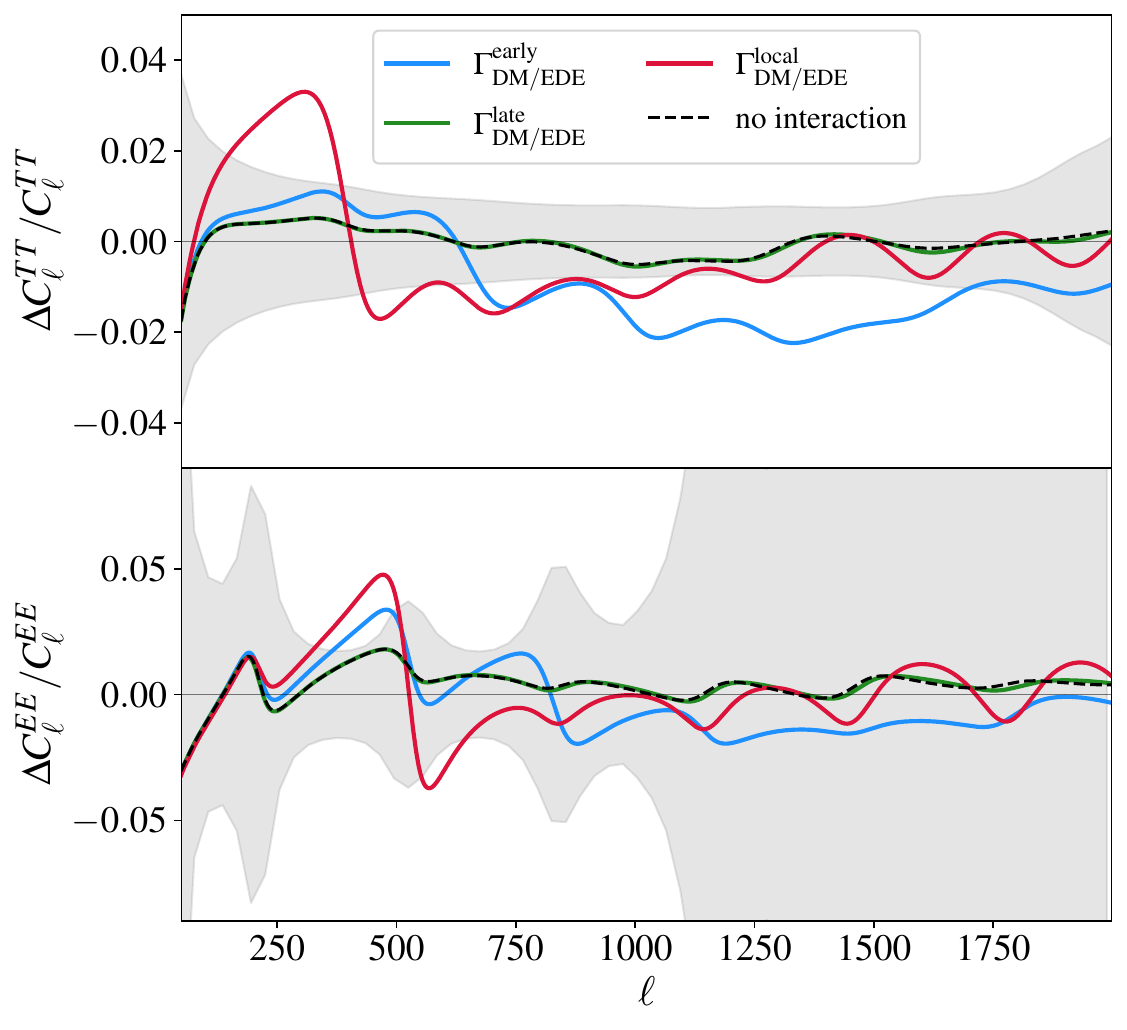}
    \caption{Residuals of the CMB TT and EE power spectra between EDE and $\Lambda$CDM, with and without the drag term. We illustrate the effect of the three parametrizations of $\Gamma$, normalized such that $\sigma_8 \simeq 0.77$. The gray bands represent the 
Planck error bars from the binned PR3 data release.}
    \label{fig:cmb}
\end{figure}

We illustrate the effect of the drag term on the matter power spectrum and CMB temperature and polarization anisotropies in Fig.~\ref{fig:pk} and Fig.~\ref{fig:cmb}, respectively, by plotting residuals with respect to $\Lambda$CDM. 
Similarly to Fig.~\ref{fig:Gamma}, we set EDE and cosmological parameters to the best-fit model of the data combination ``${\cal DH}$'', with $\beta$ adjusted by hand to give $\sigma_8 = 0.77$. 

First, in the top panel of Fig.~\ref{fig:pk}, which displays the matter power spectrum residuals at $z=0$, one can see that the drag between DM and EDE suppresses power on small scales as expected. However, the range of scales affected differs depending on whether the interaction is active prior or after $z_c$. The ``late-time'' model has a much stronger effect on the small scales than the ``local'' and ``early-time'' models, for which the interactions stop before or around the onset of matter domination. As such, the drag slows down the growth of DM perturbations much more efficiently when it is present at late-times. 

Second, the bottom panel of Fig.~\ref{fig:pk} shows that the late-time drag leaves the matter power spectrum unaffected at $z=1000$, while the effect of the drag is already imprinted at early times in the other models. Consequently,  in Fig.~\ref{fig:cmb}, one can see that the ``late-time'' drag model gives residuals in the CMB TT and EE power spectra that are identical to a model without interaction  (black and green lines are super-imposed). On the other hand, a model with drag active around $z_c$ can significantly affect the CMB. Therefore, we anticipate that models with ``early-time'' or ``local'' drag will be more easily distinguishable from regular EDE  by CMB data  than the ``late-time'' drag model.

\section{Analysis}\label{sec:analysis}

\subsection{Method and data sets} 

We perform Monte Carlo Markov Chain (MCMC) analyses, confronting the interacting EDE ($i$EDE) models with recent cosmological observations, thanks to the Metropolis-Hastings algorithm from the \texttt{MontePython-v3}\footnote{\url{https://github.com/brinckmann/montepython_public}} code~\cite{Brinckmann:2018cvx,Audren:2012wb}, which is interfaced with our modified \texttt{CLASS}~\cite{Lesgourgues:2011re,Blas:2011rf} version.\footnote{\url{https://github.com/PoulinV/AxiCLASS}}
We carry out several analyses using the following datasets:
\begin{itemize}
    \item \textbf{Planck:} The low-$\ell$ CMB temperature and polarization auto-correlations (TT, EE), and the high-$\ell$ TT, TE, EE data~\cite{Planck:2019nip}, as well as the gravitational lensing potential reconstruction from {\it Planck}~2018~\cite{Planck:2018lbu}.
    
    \item \textbf{ext-BAO:} The low-$z$ BAO data gathered from 6dFGS at $z = 0.106$ \cite{Beutler:2011hx}, SDSS DR7 at $z = 0.15$ \cite{Ross:2014qpa}.
    
    \item \textbf{EFTofBOSS:} The EFTofLSS analysis of the monopole and quadrupole of the galaxy power spectra from BOSS DR12 luminous red galaxies (LRG), cross-correlated with the reconstructed BAO parameters \cite{Gil-Marin:2015nqa}. The SDSS-III BOSS DR12 galaxy sample data and covariances are described in \cite{BOSS:2016wmc,Kitaura:2015uqa}. The measurements, obtained in \cite{Zhang:2021yna}, are from BOSS catalogs DR12 (v5)\footnote{\url{https://data.sdss.org/sas/dr12/boss/lss/}}~\cite{Reid:2015gra}.
    They are divided into four skycuts, made up of two redshift bins, namely LOWZ with $0.2<z<0.43 \  (z_{\rm eff}=0.32)$, and CMASS with $0.43<z<0.7  \ (z_{\rm eff}=0.57)$, with north and south galactic skies for each, respectively, denoted NGC and SGC.
    We use the \texttt{PyBird} code\footnote{\url{https://github.com/pierrexyz/pybird}} \cite{DAmico:2020kxu} for the theory prediction\footnote{Strictly speaking, a proper modeling of the late-time drag model would require to modify the EFTofLSS prediction, as the model does not reduce to $\Lambda$CDM at late-times. At this stage, we ignore this correction. Note that the BOSS and eBOSS data provide little constraining power in our analysis beyond the BAO measurements that are included in them and that can be applied to models beyond $\Lambda$CDM~\cite{Bernal:2020vbb}, even if it is conceivable that a proper computation of the effect of the drag at late-times would modify these results. We defer this check to future work. In addition, we note that EFTofLSS has already been used in the case of standard EDE in Refs.~\cite{DAmico:2020ods,Smith:2020rxx,Ivanov:2020ril,Simon:2022adh,Simon:2023hlp,Gsponer:2023wpm}.} as well as for the full-modeling likelihood, together with the ``West coast'' parametrization (see \textit{e.g.}, Refs~\cite{Simon:2022lde,Holm:2023laa}).
    
    \item \textbf{EFTofeBOSS:} The EFTofLSS analysis \cite{Simon:2022csv} of the monopole and quadrupole of the galaxy power spectra from eBOSS DR16 quasi-stellar objects (QSO)~\cite{eBOSS:2020yzd}. The QSO catalogs are described in \cite{Ross:2020lqz} and the covariances are built from the EZ-mocks described in \cite{Chuang:2014vfa}. There are about 343 708 quasars selected in the redshif range $0.8<z<2.2$, with $z_{\rm eff}=1.52$, divided into two skycuts, NGC and SGC~\cite{Beutler:2021eqq,Hou:2020rse}. We also use the \texttt{PyBird} code and the ``West coast'' parametrization for the eBOSS DR16 full-modeling.
    
    \item \textbf{Pantheon+:} The Pantheon+ catalog of uncalibrated luminosity distance of type Ia supernovae (SNeIa) in the range ${0.01<z<2.3}$~\cite{Brout:2022vxf}. 

    \item  $\boldsymbol{{\mathcal{S}}}$: Gaussian priors on $S_8 \equiv \sigma_8 \sqrt{\Omega_m/0.3}$ measured by the 3$\times$2pt weak lensing and galaxy clustering analyses of KiDS-1000$\times$dFLensS+BOSS, $S_8 = 0.766^{+0.020}_{-0.014}$ \cite{Heymans:2020gsg}, and DES-Y3, $S_8 = 0.776 \pm 0.017$~\cite{DES:2021wwk}. In order to calculate the Gaussian tensions and show the combined constraints on the figures, we use $S_8 = 0.771^{+0.013}_{-0.011}$, which corresponds to the weighted mean and uncertainty of the two priors.\footnote{We make this choice to emphasize the tension. The combined analysis with ``hybrid'' pipelines yield $S_8=0.790^{+0.018}_{-0.014}$ and would correspondingly reduce the level of detection of the interaction. }
    
    \item $\boldsymbol{{\mathcal{H}}}$: Gaussian prior from the late-time measurement of the absolute calibration of the SNeIa from S$H_0$ES, $M_b = -19.253 \pm 0.027$ \cite{Riess:2021jrx}, corresponding to $H_0 = (73.04\pm1.04)$ km/s/Mpc.
\end{itemize}
Our baseline dataset, denoted ``$\mathcal{D}$'', corresponds to the combination of \textit{Planck}, ext-BAO, EFTofBOSS, EFTofeBOSS and Panthon+ data.
We call ``$\mathcal{DH}$'' the combination of the baseline dataset $\mathcal{D}$ with the S$H_0$ES prior $\mathcal{H}$, whereas we denote by ``$\mathcal{DS}$'' the analyses which consider the combination of the baseline analysis $\mathcal{D}$ with the $S_8$ priors $\mathcal{S}$.
Finally, the inclusion of all the data is called $\mathcal{DHS}$.

For all runs performed, we impose wide uniform priors on the $\Lambda$CDM parameters $\{\omega_b,\omega_{\rm cdm},H_0,A_s,n_s,\tau_{\rm reio}\}$, which correspond, respectively, to the dimensionless baryon energy density, the dimensionless CDM energy density, the Hubble parameter today, the variance of curvature perturbations centered around the pivot scale $k_p = 0.05$ Mpc$^{-1}$, the scalar spectral index, and the  optical depth to reionization.
Regarding the free parameters of the EDE model, we impose a logarithmic prior on $z_c$ and uniform priors on $f_{\rm EDE}(z_c)$ and $\Theta_i$,
\begin{align*}
    3 &\le \log_{10}(z_c) \le 4, \\
    0 &\le f_{\rm EDE}(z_c) \le 0.5, \\
    0 &\le \Theta_i \le \pi. 
\end{align*}
For $\beta$, we impose the following logarithmic priors:\footnote{In practice, for the late model, we effectively run on the combination $\beta\rho_{\rm EDE, 0}$ to avoid an explicit dependence on $\rho_{\rm EDE, 0}$ in the code. As we never consider explicitly $\rho_{\rm EDE, 0}=0$, it simplifies the parameter space exploration, avoiding complicated degeneracies.}
\begin{align*}
    -25 &\le \log_{10}(\beta^{\rm early}) \le -17, \\
    -6 &\le \log_{10}(\beta^{\rm local}) \le -1, \\
    -11 &\le \log_{10}(\beta^{\rm late}) \le -5.
\end{align*}
\

In this paper, we use the \textit{Planck} neutrino treatment by considering two massless and one massive species with $m_{\nu} = 0.06$ eV \cite{Planck:2018vyg}.
We consider that our chains have converged when the Gelman-Rubin criterion $R-1<0.05$.
Finally, we extract the best-fit parameters from the procedure highlighted in the appendix of Ref.~\cite{Schoneberg:2021qvd}, and we acknowledge the use of \texttt{GetDist}~\cite{Lewis:2019xzd} to extract the probability density functions and produce our plots.

\subsection{Results for all models}

\begin{table*}
\centering
    \begin{tabular}{|l|c|c|c|c|}

        \hline
        \multicolumn{5}{|c |}{$i$EDE late} \\
        \hline
        \hline\rule{0pt}{3ex}
        & $\mathcal{D}$ &  $\mathcal{DH}$ & $\mathcal{DS}$ &  $\mathcal{DHS}$ \\
        \hline

$f_{\rm EDE}(z_c)$
	 & $< 0.035(0.020)$
	 & $0.066(0.072)\pm 0.014$
	 & $< 0.046(0.032)$ 
	 & $0.065(0.073)\pm 0.013$ 
	 \\
$\log_{10}(z_c)$
	 & $> 3.30(3.95)$
	 & $3.794(3.818)^{+0.070}_{-0.088}$
 & $> 3.29(3.88)$ 
	 & $3.778(3.818)\pm 0.076$
	 \\
	 
$\Theta_i$
	 & unconstrained$(3.10)$
	 & $> 1.91(2.85)$
	 & unconstrained$(3.03)$
	 & $> 1.49(2.85)$ 
	 \\

$\log_{10}[\Gamma/H_0]$
	 & $< 2.84(0.52)$
	 & $< 2.24(0.74)$
	 & $2.1(1.97)^{+0.52}_{-0.68}$
	 & $1.50(1.51)^{+0.32}_{-0.27}$  
	 \\
	 
\hline
$H_0$
	 & $68.32(68.65)^{+0.48}_{-0.77}$
	 & $71.49(71.80)\pm 0.74$
	 & $68.71(69.31)^{+0.50}_{-0.86}$ 
	 & $71.48(71.85)\pm 0.70$ 
	 \\	 
$\omega{}_{\rm idm }$
	 & $0.1215(0.1226)^{+0.0014}_{-0.0030}$
	 & $0.1307(0.1320)\pm 0.0032$
	 & $0.1225(0.1225)^{+0.0017}_{-0.0032}$ 
	 & $0.1304(0.1321)\pm 0.0029$ 
	 \\
$10^{2}\omega{}_{b }$
	 & $2.245(2.256)^{+0.015}_{-0.017}$
	 & $2.292(2.294)\pm 0.019$
	 & $2.252(2.262)^{+0.017}_{-0.019}$ 
	 & $2.290(2.295)\pm 0.019$ 
	 \\
$10^{9}A_{s }$
	 & $2.112(2.117)^{+0.030}_{-0.034}$
	 & $2.152(2.146)\pm 0.034$
	 & $2.113(2.120)^{+0.028}_{-0.033}$ 
	 & $2.155(2.150)^{+0.030}_{-0.038}$ 
	 \\
$n_{s }$
	 & $0.9677(0.9721)^{+0.0043}_{-0.0060}$
	 & $0.9887(0.9914)\pm 0.0059$
	 & $0.9709(0.9760)^{+0.0045}_{-0.0067}$ 
	 & $0.9884(0.9916)\pm 0.0058$ 
	 \\
$\tau{}_{\rm reio }$
	 & $0.0570(0.0575)\pm 0.0078$
	 & $0.0606(0.0595)\pm 0.0076$
	 & $0.0567(0.0568)^{+0.0065}_{-0.0075}$ 
	 & $0.0614(0.0602)^{+0.0071}_{-0.0088}$ 
	 \\
	 \hline
$S_8$
	 & $0.800(0.821)^{+0.022}_{-0.014}$
	 & $0.785(0.813)^{+0.033}_{-0.018}$
	 & $0.781(0.779)\pm 0.012$ 
	 & $0.776(0.777)^{+0.011}_{-0.013}$ 
	 \\	 
$\Omega{}_{m }$
	 & $0.3099(0.3094)^{+0.0057}_{-0.0051}$
	 & $0.3018(0.3017)^{+0.0045}_{-0.0051}$
	 & $0.3085(0.3082)\pm 0.0053$ 
	 & $0.3014(0.3015)^{+0.0046}_{-0.0053}$ 
	 \\

\hline
\end{tabular}
\caption{Mean (best-fit) $\pm 1\sigma$ (or $2\sigma$ for one-sided bounds) of reconstructed parameters of the $i$EDE late model confronted to various datasets.}
\label{tab:cosmoparam_iEDE}
\end{table*}

\begin{table*}[]
    \centering
    \scalebox{0.96}{
    \begin{tabular}{|c|c c c c| c c c c|c c c c|c c c c|}
         \hline
         & \multicolumn{4}{c | }{EDE}& \multicolumn{4}{c |}{$i$EDE early} & \multicolumn{4}{c |}{$i$EDE local} & \multicolumn{4}{c |}{$i$EDE late}\\
         \hline
         \hline\rule{0pt}{3ex}
        & $\mathcal{D}$  & $\mathcal{DH}$ & $\mathcal{DS}$& $\mathcal{DHS}$& $\mathcal{D}$  & $\mathcal{DH}$ & $\mathcal{DS}$ & $\mathcal{DHS}$ & $\mathcal{D}$  & $\mathcal{DH}$ & $\mathcal{DS}$ & $\mathcal{DHS}$ & $\mathcal{D}$  & $\mathcal{DH}$ & $\mathcal{DS}$ & $\mathcal{DHS}$\\
        \hline \rule{0pt}{3px}
       $Q^{H_0}\equiv Q_\mathrm{DMAP}$ tension $H_0$ & $3.2\sigma$ & -- & $3.7\sigma$ & -- 
       & $3.3\sigma$ & -- & $3.6\sigma$ & -- 
       & $3.3\sigma$ & -- & $4.2\sigma$ & -- 
       & $3.3\sigma$ & -- & $3.1\sigma$ & --\\
       $Q^{S_8}\equiv Q_\mathrm{DMAP}$ tension $S_8$ & $3.5\sigma$ & $3.9\sigma$ & -- & -- 
       & $2.9\sigma$ & $3.3\sigma$ & -- & -- 
       & $3.0\sigma$ & $3.9\sigma$ & -- & -- 
       & $1.7\sigma$ & $1.2\sigma$ & -- & -- \\ 
       \hline
        $\Delta \chi^2$ & $ -3.2 $ & $-30.1 $ & $-0.7$ & $-19.6$ 
        & $-3.3$ & $-30.1$ & $-4.9$ & $-24.4$ 
        & $-3.8$ & $-30.1$ & $-4.9$ & $-19.8$  
        & $-3.7$ & $-30.3$ & $-10.6$ & $-33.9$  \\
        $\Delta$AIC & $+2.8$ & $-24.1$ & $+5.3$ & $-13.6$ 
        & $+4.7$ & $-22.1$ & $+ 3.1$ & $-16.4$  
        & $+4.2$ & $-22.1$ & $+3.1$ & $-11.8$  
        & $+4.3$ & $-22.3$ & $-2.6$ & $-25.9 $  \\
        \hline
    \end{tabular}}
    \caption{$Q_\mathrm{DMAP}$ tensions, differences in the minimized effective $\chi^2$, as well as the associated $\Delta$AIC$=\Delta\chi^2+2\times\Delta N_p$ where $\Delta N_p$ is the difference in the number of free parameters between models, for various combinations of data for the EDE and $i$EDE models.}
    \label{tab:tensions}
\end{table*}

\begin{figure*}
    \centering
    \includegraphics[width=1.5\columnwidth]{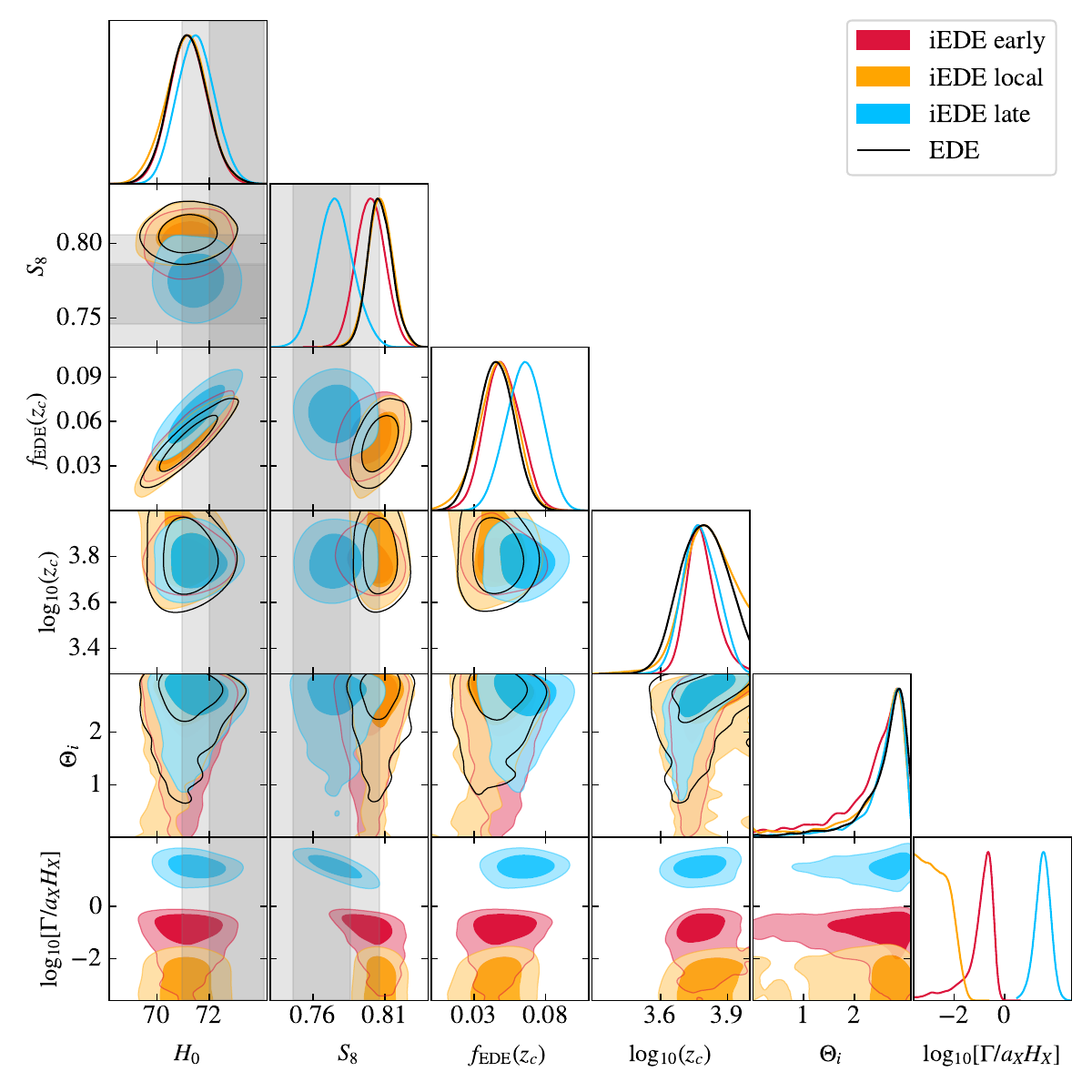}
    \caption{2D posterior distributions reconstructed from the $\mathcal{DHS}$ dataset for the standard EDE model and the three interacting EDE scenarios. The gray bands show the $H_0$ and $S_8$ priors defined in Sec.~\ref{sec:analysis}. Note that $\log_{10}[\Gamma / a_X H_X]$ is evaluated at $a_X=a_c$ for the early and local interaction scenarios and is evaluated at $a_X=a_0$ for the late interaction scenario.}
    \label{fig:general_comparison}
\end{figure*}

\begin{figure*}
    \centering
    \includegraphics[width=1.5\columnwidth]{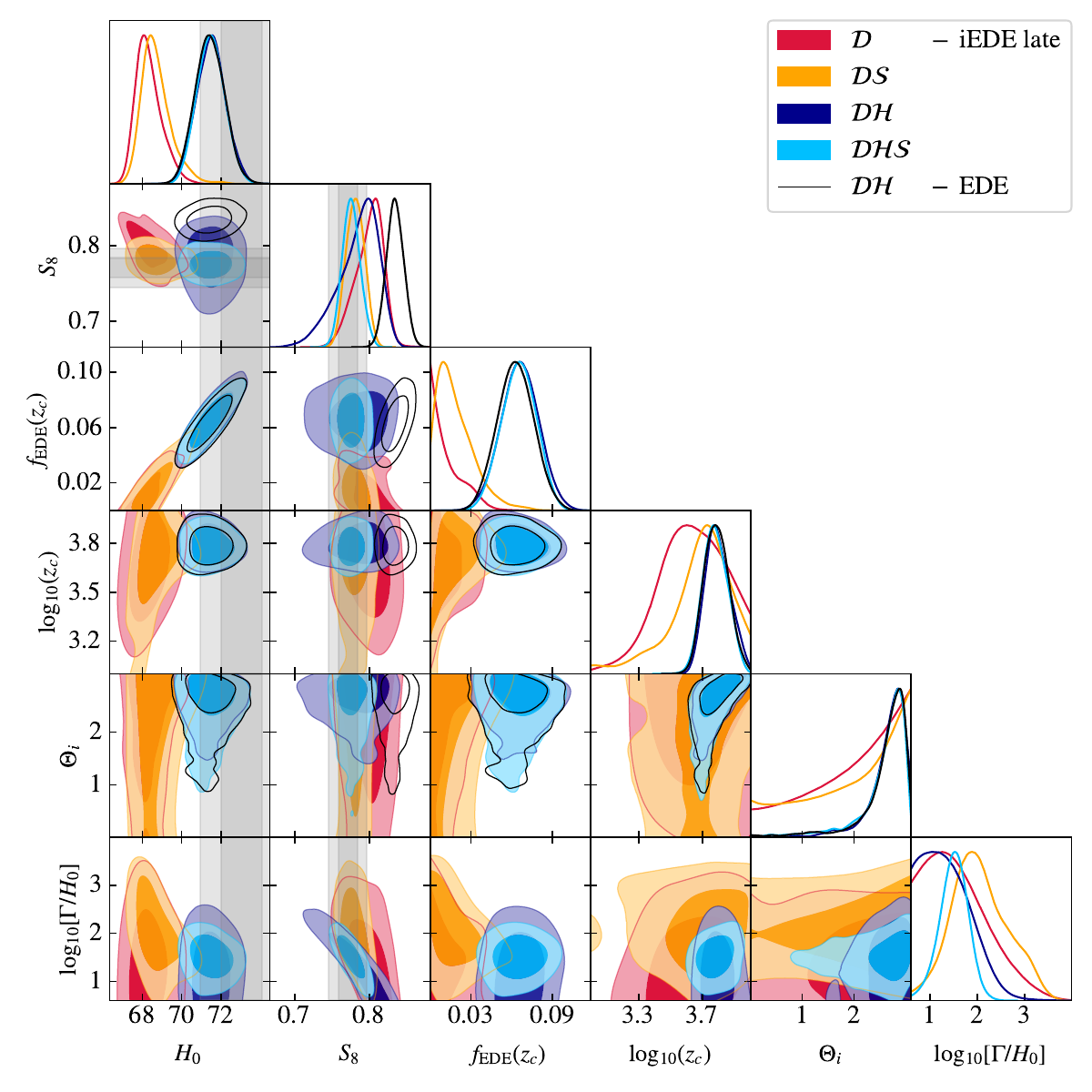}
    \caption{2D posterior distributions reconstructed from the $\mathcal{D}$, $\mathcal{DH}$, $\mathcal{DS}$ and $\mathcal{DHS}$ datasets for the late-interaction EDE scenario. For comparison, we also display the standard EDE 2D posterior distributions reconstructed from the $\mathcal{DH}$ dataset. The gray bands show the $H_0$ and $S_8$ priors defined in Sec.~\ref{sec:analysis}.}
    \label{fig:iEDE_late}
\end{figure*}

In this section, we discuss the ability of the three $i$EDE scenarios to resolve the $S_8$ tension on top of the $H_0$ tension. The cosmological constraints of the standard EDE model as well as the three $i$EDE scenarios are displayed in Tabs.~\ref{tab:cosmoparam_iEDE}, ~\ref{tab:cosmoparam} and ~\ref{tab:cosmoparams_early_local} for the $\mathcal{D}$, $\mathcal{DH}$, $\mathcal{DS}$ and $\mathcal{DHS}$ datasets. In addition, the $Q_{\rm DMAP}$ tensions for $S_8$ and $H_0$, as well as the $\Delta\chi^2_{\rm min}$ and the associated $\Delta$AIC with respect to $\Lambda$CDM, are summarized in Tab.~\ref{tab:tensions}. In the latter table, we define 
\begin{eqnarray}
       Q^{H_0}_{\mathcal{DH}} & = & \sqrt{\chi_{\rm min}^2(\mathcal{DH})-\chi_{\rm min}^2(\mathcal{D})} \, ,\\
       Q^{H_0}_{\mathcal{DHS}}& = & \sqrt{\chi_{\rm min}^2(\mathcal{DHS})-\chi_{\rm min}^2(\mathcal{DS})} \, ,
\end{eqnarray}
as well as 
\begin{eqnarray}
    Q^{S_8}_{\mathcal{DS}}& = &\sqrt{\chi_{\rm min}^2(\mathcal{DS})-\chi_{\rm min}^2(\mathcal{D})} \, ,\\
       Q^{S_8}_{\mathcal{DHS}}& = &\sqrt{\chi_{\rm min}^2(\mathcal{DHS})-\chi_{\rm min}^2(\mathcal{DH})} \, .
\end{eqnarray}
Finally, in Tab.~\ref{tab:chi2} of App.~\ref{app:chi2}, we show the $\chi^2_{\rm min}$ associated with each likelihood for the different models and combination of datasets considered in this work.

Our main results are summarised in Fig.~\ref{fig:general_comparison}, which shows the 2D posterior distributions reconstructed from the $\mathcal{DHS}$ dataset for the different scenarios studied in this paper. In addition, we display, for the $i$EDE late scenario, the 2D posterior distributions reconstructed from the $\mathcal{D}$, $\mathcal{DH}$, $\mathcal{DS}$ and $\mathcal{DHS}$ datasets in Fig.~\ref{fig:iEDE_late}, while in App.~\ref{app:early_and_local} we show the same figure for the $i$EDE early and $i$EDE local models.
We present here our main results based on these tables and figures:
\begin{itemize}
    \item Firstly, when $S_8$ priors are excluded from the analyses, all models lead to a similar alleviation of the $H_0$ tension with $Q^{H_0}_{\mathcal{DH}} = 3.2 - 3.3 \sigma$ (see Tab.~\ref{tab:tensions}) compared to $\Lambda$CDM, where $Q^{H_0}_{\mathcal{DH}} = 6.1 \sigma$. This indicates that the presence of the drag does not further help reducing the $H_0$ tension. 
    \item Secondly, one can see that the discrepancy in $S_8$ measurements can be alleviated in the $i$EDE late model, with $Q^{S_8}_{\mathcal{DS}} = 1.7 \sigma$ and $Q^{S_8}_{\mathcal{DHS}} = 1.2 \sigma$. Consequently, one finds a detection of the coupling strength $\beta$ (as required to lower $S_8$), with a preference for non-zero $\beta$ at the 3.8$\sigma$ level ($\Delta\chi^2=14.3$ with respect to EDE, for 1 extra degree of freedom).  On the other hand, the early and local $i$EDE models, as well as regular EDE, cannot achieve low-$S_8$, with $Q^{S_8}_{\mathcal{DS}} = 2.9 - 3.5 \sigma$ and $Q^{S_8}_{\mathcal{DHS}} = 3.3 - 3.9 \sigma$.  The early drag model shows a mild level of detection of $\beta$ with the $\mathcal{DS}$ and $\mathcal{DHS}$  dataset ($2\sigma$ level), but it is not enough to reduce significantly $S_8$. The local model also shows a similar mild preference for non-zero $\beta$ in the $\mathcal{DS}$ case, but it becomes indistinguishable from regular EDE in the $\mathcal{DHS}$ case (see Fig.~\ref{fig:iEDE_early}).
    
   \item  Finally, the $H_0$ tension increases for the $\mathcal{DHS}$ dataset (compared to the $\mathcal{DH}$ dataset) for the EDE, $i$EDE early and $i$EDE local models, as a consequence of the fact that a too large $H_0$ is disfavored when including the low-$S_8$ priors. However, in the late $i$EDE model, the $H_0$ tension is slightly lower than previously, with $Q^{H_0}_{\mathcal{DHS}} = 3.1 \sigma$.

\end{itemize}

In conclusion, this confirms the intuition from Sec.~\ref{sec:impact_CMB_LSS} that the early and local models are more strongly constrained given the impact on the CMB power spectra. This also shows that, while a drag in the DM fluid active at late-times can easily lower $S_8$ without spoiling the fit to other datasets, the drag cannot occur at the same time as the EDE energy injection. We explore the reasons for those constraints further in the next section.

\subsection{Anatomy of the interactions}

In order to understand further why only the late $i$EDE model achieves a reduction of the $S_8$ and $H_0$ tensions simultaneously, we compare results of the $\mathcal{DS}$ and $\mathcal{DH}$ analyses in Fig.~\ref{fig:iEDE_late}. First and foremost, one can see a clear degeneracy between $\log_{10}[\Gamma/H_0]$ and $S_8$ in the $\mathcal{DH}$ analysis (dark blue), with a clear overlap of the posteriors with the direct measurements from KiDS and DES (represented in grey). 
In addition, in the $\mathcal{DS}$ analysis, one can see that the late $i$EDE model is also favored with respect to both $\Lambda$CDM and regular EDE models, with a detection of $\log_{10}[\Gamma/H_0] = 2.1^{+0.52}_{-0.68}$ with $f_{\rm EDE} <  0.046 $. Note that the absence of a lower limit on $f_{\rm EDE}$ in that case is due to the choice of a linear prior on this parameter. Running with a logarithmic prior on $f_{\rm EDE}$, which emphasizes small values of this parameter, provides a detection of $f_{\rm EDE}$ as shown in the bottom panel of Fig.~\ref{fig:iEDE_DS_log10fede} (see below).

Of particular interest is the plane $\log_{10}(z_c)$ vs $S_8$ in Fig.~\ref{fig:iEDE_late}: it shows that the $\mathcal{DS}$ and $\mathcal{DH}$ datasets prefer the same range of $z_c$ for the EDE to be active. Because the drag is only active at late-times, the degeneracy between $f_{\rm EDE}$ and $H_0$ that exists for a narrow range of $z_c$ can be exploited without spoiling the fit. 
As there is a consistency in the range of $z_c$ favored in both datasets, their combination allows us to resolve both tensions simultaneously. 

Comparing with the early and local models of $i$EDE, shown in Fig.~\ref{fig:iEDE_early} and Fig.~\ref{fig:iEDE_local} of App.~\ref{app:early_and_local}, one can see that the results are very different. In the $\mathcal{DS}$ analysis, only very small fraction of $f_{\rm EDE}(z_c)$ is allowed. However, we note that a lower $S_8$ can be reached with a corresponding increase $\log_{10}(z_c)$ compared to the $\mathcal{DH}$ dataset. This suggests that it could be possible to resolve the $S_8$ tension only at much larger values of $z_c$ and much smaller fraction of $f_{\rm EDE}(z_c)$ than that favored by the $\mathcal{DH}$ analysis. This indicates that the parameter space region favored in both analyses are mutually inconsistent, and that an EDE-DM drag active prior to recombination cannot resolve both tensions. 

To confirm this, we present in Fig.~\ref{fig:iEDE_DS_log10fede} the results of analyses of the three $i$EDE scenarios for the $\mathcal{DS}$ dataset with larger prior on $\log_{10}(z_c)$ and a log-prior on the EDE fraction $f_{\rm EDE}(z_c)$, to emphasize smaller EDE fractions. 
Interestingly, in the late iEDE scenario,  extending the $z_c$ prior reveals that the data actually favor a drag occurring after the era of recombination ($\log_{10}(z_c)<3$), which highlights that the drag is preferably not active while the EDE boosts the pre-recombination era expansion rate. 
In addition, both local and early drag models also manage to lower $S_8$ for small fraction $f_{\rm EDE}(z_c)\sim 10^{-4} - 10^{-2}$, as expected. Although the EDE contributes before recombination, for such small fraction, the impact on the sound horizon and $H_0$ is negligible. One cannot find a simultaneous solution to both tensions in those models, confirming the results with a linear prior on $f_{\rm EDE}(z_c)$.

\begin{figure*}
    \centering
    \includegraphics[width=1.5\columnwidth]{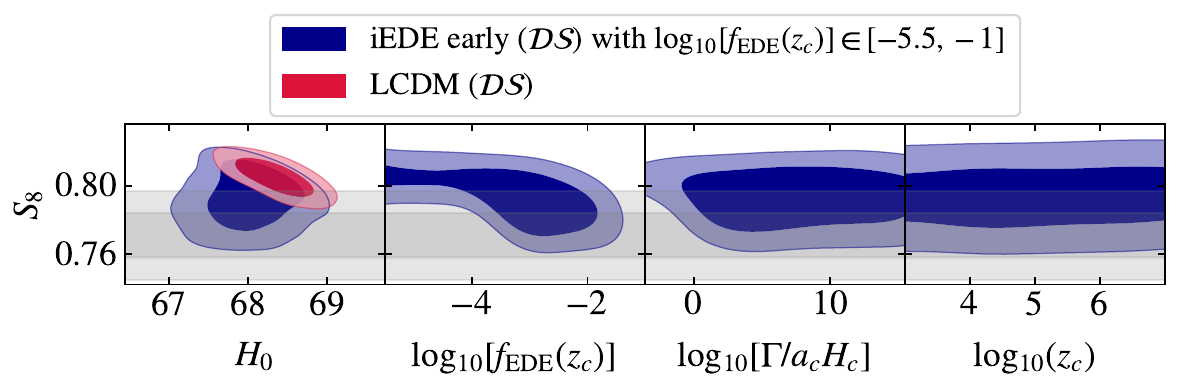}
    \includegraphics[width=1.5\columnwidth]{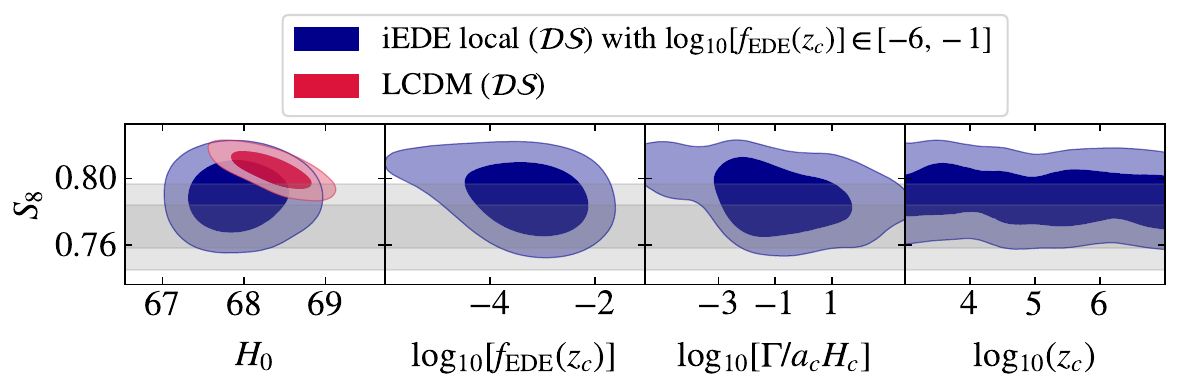}
    \includegraphics[width=1.5\columnwidth]{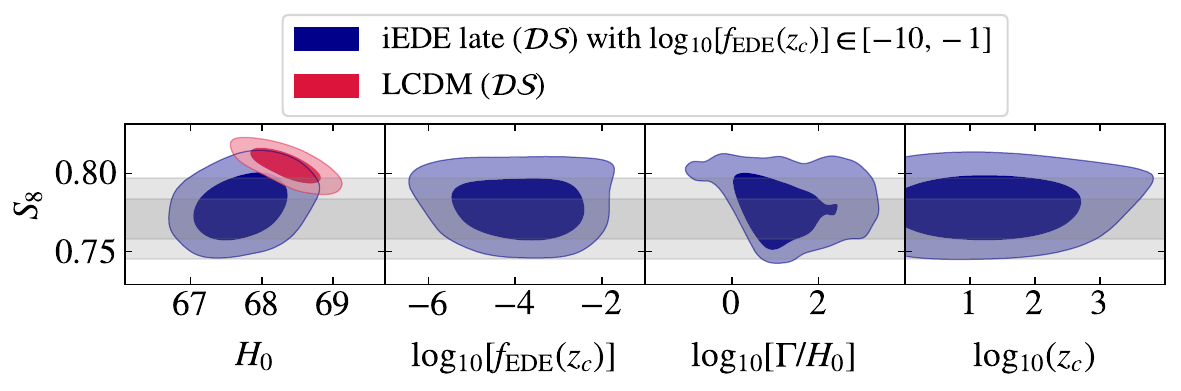}
    \caption{2D posterior distributions reconstructed from the $\mathcal{DS}$ dataset for the early, local, and late iEDE scenarios with a logarithmic prior on $f_{\rm EDE}$. For comparison, we also display the $\Lambda$CDM constraints on $\{H_0,S_8\}$, while the gray bands show the $S_8$ priors defined in Sec.~\ref{sec:analysis}.
    In the early analysis (\textit{top panel}), we vary $\log_{10}{\beta} \in [-25,-13]$ and $\log_{10}{z_c} \in [3,7]$, in the local analysis (\textit{middle panel}), we vary $\log_{10}{\beta} \in [-8,3]$ and $\log_{10}{z_c} \in [3,7]$,
    and in the late analysis (\textit{bottom panel}), we vary $\log_{10}{\beta} \in [-20,-5]$ and $\log_{10}{z_c} \in [0,7]$.
    }
    \label{fig:iEDE_DS_log10fede}
\end{figure*}

\section{Conclusions}
\label{sec:conclusions}
The origin of cosmic tensions in recent years remains an open question extensively explored. In fact, explaining simultaneously the $H_0$ and $S_8$ tensions has proven challenging because, at face value, the S$H_0$ES calibration of the cosmic distance ladder implies a larger $\omega_{\rm cdm}$, which in turn leads to an earlier matter domination and a larger $S_8$ \cite{Jedamzik:2020krr,Poulin:2024ken}.
Hence, models that can explain the $H_0$ tension, such as the EDE models, typically lead to an increase in the $S_8$ tension. 
However, it has been shown that the inclusion of a drag term in the DM Euler equation can reduce $S_8$, as is done for instance in models of DM-DR interactions, or models of (late-time) DE-DM drag.
In this paper, we have investigated whether a drag with DM induced by an EDE component can simultaneously alleviate the $H_0$ and $S_8$ tensions.

Building upon the axion-like EDE model, we have introduced a phenomenological drag model between EDE and DM with three different parametrizations of the drag rate, based solely on ensuring energy-momentum conservation, providing thereby flexibility in the modelling.  We have paid particular attention to identifying {\it when} the drag can be introduced in an optimal way.
Our results can be summarized as follows:

\begin{itemize}
    \item We have found that only the ``late'' drag model (where the drag is relevant post-recombination) allows a simultaneous alleviation of the $H_0$ and $S_8$ tensions, with a preference for a non-zero drag at the $3.8\sigma$ level and no residual $S_8$ tension. 
    \item The ``early'' drag model (where the drag is relevant at all times until $z_c$) shows a $2\sigma$ preference for a drag when the $S_8$ prior is included (${\cal DS}$ and ${\cal DHS}$ analyses), but the $S_8$ tension remains at the $2.9\sigma$ level. 
     \item The ``local'' drag model (where the drag is relevant only around $z_c$) is also mildly favored over regular EDE in the ${\cal DS}$ analysis but cannot alleviate the $H_0$ and $S_8$ simultaneously.
    \item All models yield a similar level of alleviation of the $H_0$ tension to the $3.2-3.3\sigma$ (down from $6.1\sigma$ in $\Lambda$CDM). The presence of the drag does not further relax the $H_0$ tension in a significant way. It only contributes to reducing the $S_8$ tension.
    
\end{itemize}

Our results suggest that the drag cannot be large at the time at which EDE represents a sizeable fraction of the energy density. This is a challenge to this model, given that it appears at face value better motivated for the drag to be important precisely at the time of the EDE energy injection.
As we show in App.~\ref{sec:discussion}, introducing a coupling between a scalar field and dark matter at the Lagrangian level introduces additional effects not captured by our simple parametrization.  Our results, while providing a first insight into the exploration of the EDE-DM drag resolution of the cosmic tensions, should thus be understood with some caveats.
It will be interesting to extend this work by considering couplings directly at the Lagrangian level in future work to firmly confirm these results.

We note that there are examples of other models which introduce additional energy density pre-recombination, such as `Wess-Zumino dark radiation', that can both address the Hubble tension and, through a coupling with dark matter, reduce the value of $S_8$ \cite{Joseph:2022jsf,Allali:2023zbi,Buen-Abad:2023uva,Schoneberg:2023rnx}. This highlights the fact that the specific dynamics determine whether a model can address both tensions, motivating a more systematic approach to determine the phenomenological requirements for such models. 

\begin{acknowledgements}

The authors acknowledge the use of computational resources from the Excellence Initiative of Aix-Marseille University (A*MIDEX) of the “Investissements d’Avenir” programme. 
This project has received support from the European Union’s Horizon 2020 research and innovation program under the Marie Skodowska-Curie grant agreement No 860881-HIDDeN.  This project has also received funding from the European Research Council (ERC) under the
European Union’s HORIZON-ERC-2022 (Grant agreement No. 101076865).  JLB acknowledges funding from the Ramón y Cajal Grant RYC2021-033191-I, financed by MCIN/AEI/10.13039/501100011033 and by
the European Union “NextGenerationEU”/PRTR, as well as the project UC-LIME (PID2022-140670NA-I00), financed by MCIN/AEI/ 10.13039/501100011033/FEDER, UE. EDK acknowledges  joint support from the U.S.-Israel Bi-national Science Foundation (BSF, grant No. 2022743) and  the U.S. National Science Foundation (NSF, grant No. 2307354), as well as support from the ISF-NSFC joint research program (grant No. 3156/23). 
\end{acknowledgements}

\appendix

\section{A pure moment transfer between scalar field early dark energy and dark matter}\label{sec:discussion}

Motivating the phenomenological model we outlined above with a theory derived from first principles adds an intriguing aspect to our study. As previously discussed, the simplest coupling can be described by the  Lagrangian given by Eq.~\eqref{eq:lagrangian}, which we recall here~\cite{Pourtsidou:2016ico,Pourtsidou:2013nha}
\begin{equation}
    \mathcal{L}=\frac{1}{2}\nabla_\mu\phi\nabla^\mu\phi+\beta \left(u^\mu\nabla_\mu\phi\right)^2 - V(\phi) \, ,
\end{equation}
where $\phi$ is a quintessence scalar field, $u^\mu$ is the four-velocity of the fluid, $V(\phi)$ is the potential, and $\beta$ is a coupling constant. We limit ourselves to $\beta<1/2$ to avoid a negative kinetic term. In order to account for the perturbations, we express $\phi$ in terms of the background and perturbed contributions, $\phi=\phi_0 + \phi_1$. In these terms, the energy density and pressure of the field are~\cite{Pourtsidou:2016ico}
\begin{align}
    \overline{\rho}_{\phi}&=\left(\frac{1}{2}-\beta\right)\frac{\dot{\phi}_{0}^{2}}{a^{2}}+V\left(\phi\right) \, ,\\\overline{P}_{\phi}&=\overline{\rho}_{\phi}-2V\left(\phi\right) \, ,
\end{align}
at the background level, and
\begin{align}
    \delta\rho_{\phi}&=\left(1-2\beta\right)\frac{\dot{\phi}_{0}\dot{\phi}_{1}}{a^{2}}+V_{,\phi}\phi_{1} \, ,\\
    \delta P_{\phi}&=\delta\rho_{\phi}-2V_{,\phi}\phi_{1} \, ,
\end{align}
at the linearly perturbed level, where subscripts `,$\phi$' refer to derivatives with respect to the scalar field.
The equation of motion for the scalar field at the background level is 
\begin{equation}
    \left(1-2\beta\right)\left(\ddot{\phi}_{0}+2\frac{\dot{a}}{a}\dot{\phi}_{0}\right)+a^{2}V_{,\phi}=0 \, ,
\end{equation}
while at the perturbed level we have (following the conventions in Ref.~\cite{Ma:1995ey} for the metric and fluid perturbations in \textit{synchronous} gauge)
\begin{align}
    \left(1-2\beta\right)&\left(\ddot{\phi}_{1}+ 2\frac{\dot{a}}{a}\dot{\phi}_{1}\right)+ 
    \left( k^{2} +a^{2} V_{,\phi\phi}\right)\phi_{1}+\nonumber\\
    &\frac{1}{2}\left(1-2\beta\right)\dot{\phi}_{0}\dot{h}-2\beta\dot{\phi}_{0}\theta_{\rm DM}=0 \, ,
\end{align}
where $h$ is the trace of the metric perturbations, and 
\begin{equation}
    \theta_{\phi}=\frac{1}{1-2\beta}\left(k^{2}\frac{\phi_{1}}{\dot{\phi}_{0}}-2\beta \theta_{\rm DM}\right) \, ,
\end{equation}
is the divergence of the fluid velocity.\footnote{Note that this convention differs than the one in Refs.~\cite{Pourtsidou:2016ico,Pourtsidou:2013nha} by a factor of $k^2$.} 
The Euler equation for the CDM now reads
\begin{equation}
    \dot{\theta}_{\rm DM}+\frac{\dot{a}}{a}\theta_{\rm DM}=-2\beta\frac{\left(\frac{\ddot{\varphi}_{0}}{a}+2\frac{\dot{a}}{a}\frac{\dot{\varphi}_{0}}{a}\right)\varphi_{1}+\frac{\dot{\varphi}_{0}}{a}\dot{\varphi}_{1}}{a\left(\overline{\rho}_{\rm DM}-2\beta\frac{\dot{\varphi}_{0}^{2}}{a^{2}}\right)}k^{2} \, .
\end{equation}
Writing the Euler equations for both species in terms of the field fluid quantities, we find
\begin{align}\label{eq:EulerGen}
    \dot{\theta}_{\rm DM}=&-\frac{\dot{a}}{a}\theta_{\rm DM}-G \, ,\\
    \dot{\theta}_{\phi}=&2\frac{\dot{a}}{a}\theta_{\phi}+\frac{1}{1-2\beta}\frac{\delta\rho_{\phi}}{\overline{\rho}_{\phi}+\overline{P}_{\phi}}k^{2}\nonumber\\
    &+\frac{2\beta}{1-2\beta}\left(3\frac{\dot{a}}{a}\theta_{\rm DM}+G\right) \, ,\nonumber
\end{align}
where
\begin{equation}
    G\equiv \frac{2\beta}{1-2\beta}\frac{\frac{\delta\rho_{\phi}}{\overline{\rho}_{\phi}+\overline{P}_{\phi}}k^{2}+3\frac{\dot{a}}{a}\left(1-c_{\phi}^{2}\right)\left[\theta_{\phi}+2\beta\left(\theta_{\rm DM}-\theta_{\phi}\right)\right]}{\frac{\overline{\rho}_{\rm DM}}{\overline{\rho}_{\phi}+\overline{P}_{\phi}}-\frac{2\beta}{1-2\beta}} \, ,
\end{equation}
and $c_\phi^2\equiv \dot{\overline{P}}_\phi/\dot{\overline{\rho}}_\phi$ is the adiabatic sound speed.
In the limit of weak momentum coupling, $\left|\beta\right|\ll 1$, Eq.~\eqref{eq:EulerGen} simplify and now read
\begin{align}\label{eq:EulerSimp}
    \dot{\theta}_{\rm DM}\approx&-\frac{\dot{a}}{a}\theta_{\rm DM}-2\beta\left(\frac{\delta\rho_{\phi}}{\overline{\rho}_{\rm DM}}k^{2}+3\frac{\dot{a}}{a}\frac{\overline{\rho}_{\phi}+\overline{P}_{\phi}}{\overline{\rho}_{\rm DM}}\left(1-c_{\phi}^{2}\right)\theta_{\phi}\right) ,\nonumber\\
    \dot{\theta}_{\phi}\approx& 2\frac{\dot{a}}{a}\theta_{\phi}+\frac{\delta\rho_{\phi}}{\overline{\rho}_{\phi}+\overline{P}_{\phi}}k^{2}+6\beta\frac{\dot{a}}{a}\theta_{c} \, .
\end{align}
While the DM continuity equation is left unaffected, by looking at the full DM evolution in Eq.~\eqref{eq:EulerGen}, or even the  weak coupling limit \eqref{eq:EulerSimp}, it is clear that the phenomenological model is missing some of the features of a theory derived from first principles. In addition, the modification to the KG equation suggests that the dynamics of the field can also be affected by this coupling. It will thus be interesting to go beyond the phenomenological approach and test this model explicitly (see \textit{eg}, Ref.~\cite{Liu:2023haw}), although we expect our approach to lead to conservative constraints as the effect of the coupling is limited to a drag term.

\section{Review of the fluid model of EDE}
\label{app:ede_fluid}
The fluid EDE model (see Ref.~\cite{Poulin:2023lkg}) has an energy density that follows 
\begin{equation}
\rho_{\rm EDE}(a) = \rho_{\rm EDE,0} e^{3\int_a^1 [1+w_{\rm EDE}(a)]da/a} \,,
\end{equation}
where $w_{\rm EDE}(a)$ is the EDE equation of state defined as
\begin{equation}
\label{eq:w_EDE_pheno}
w_{\rm EDE}(a) = \frac{1+w_f}{1+(a_c/a)^{3(1+w_f)}} -1\,,
\end{equation}
such that the EDE fluid has $w_{\rm EDE} \to -1$ when $a \ll a_c$, and $w_{\rm EDE} \to w_f$ when $a \gg a_c$. This parametrization captures the background dynamics of an axion-like scalar field $\phi$ with a potential $V(\phi)\propto (1-\cos(\phi/f))^n$ and with a final equation of state $w_f=(n-1)/(n+1)$ \cite{Poulin:2018cxd}. We fix $n=3$ for simplicity, since data have been shown to be not highly sensitive to this parameter as long as $2\lesssim n\lesssim 5$. 
The dynamics of perturbations in the fluid approximation is dictated by the effective sound-speed $c_s^2$, that can be approximated, for an oscillating scalar field with a potential $\phi^{2n}$, as \cite{Poulin:2018cxd}
\begin{equation}
\label{eq:cs2_axEDE}
c_{s}^2(k,a) = \frac{2 a^2(n-1) \varpi^2+k^2}{2 a^2 (n+1) \varpi^2 + k^2} \,,
\end{equation}
where $\varpi$ is the angular frequency of the oscillating background field, well-approximated by \cite{Johnson:2008se,Poulin:2018dzj,Smith:2019ihp}
\begin{eqnarray}
\varpi(a) &\simeq& m\frac{\sqrt{\pi} \Gamma(\frac{1+n}{2n})}{\Gamma\left(1+\frac{1}{2n}\right)}2^{-(1+n)/2}  \theta^{n-1}_{\rm env}(a)\,,
\label{eq:omega}\\
&\simeq& 3 H(z_c)\frac{\sqrt{\pi} \Gamma(\frac{1+n}{2n})}{\Gamma \left(1+\frac{1}{2n}\right)}2^{-(1+n)/2} \frac{\theta^{n-1}_{\rm env}(a)}{{\sqrt{|E_{n,\theta \theta}(\theta_i)|}}}\,.
\nonumber
\end{eqnarray}
In this equation, we have written the scalar field potential as $V_n(\phi) = m^2 f^2 E_n(\theta = \phi/f)$, where $m$ and $f$ are respectively the axion mass and decay constant.
In addition, $\Gamma(x)$ is the Euler Gamma function corresponding to the envelope of the background field once it is oscillating, $\theta_{\rm env}\equiv \phi_{\rm env}/f$, where
\begin{equation}
\phi_{\rm env}(a)=\phi_c\left(\frac{a_c}{a}\right)^{3/(n+1)}\,.\label{eq:phi_env}
\end{equation}

As usual for the axion-like EDE model,  $f_{\rm EDE}(z_c)$, $\log_{10}(z_c)$ and $\Theta_i$ the three free parameters describing the EDE dynamics. 

\section{Results for the local and early interaction models} 
\label{app:early_and_local}

\begin{figure*}
    \centering
    \includegraphics[width=1.5\columnwidth]{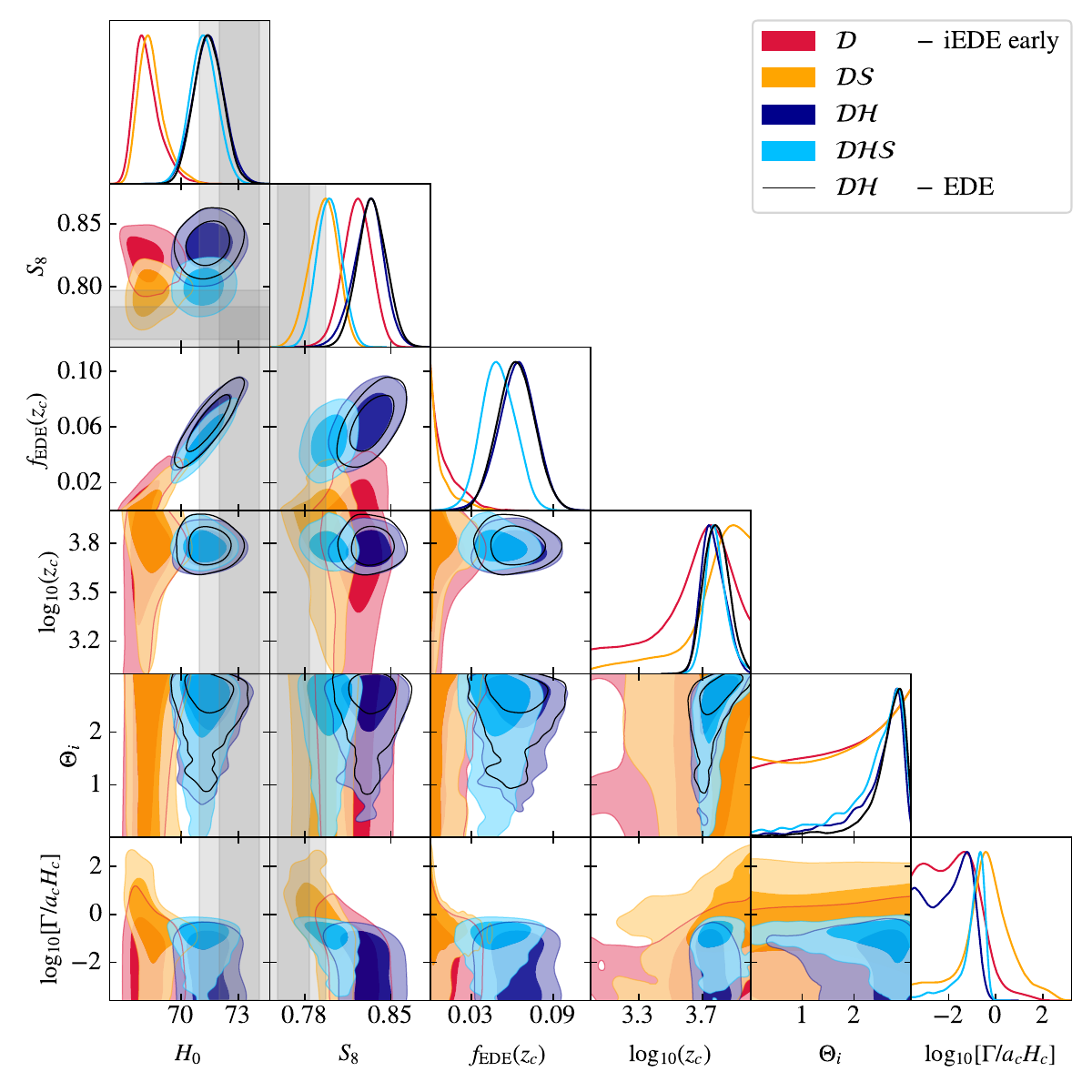}
    \caption{2D posterior distributions reconstructed from the $\mathcal{D}$, $\mathcal{DH}$, $\mathcal{DS}$ and $\mathcal{DHS}$ datasets for the early-interaction EDE scenario. For comparison, we also display the standard EDE 2D posterior distributions reconstructed from the $\mathcal{DH}$ dataset. The gray bands show the $H_0$ and $S_8$ priors defined in Sec.~\ref{sec:analysis}.}
    \label{fig:iEDE_early}
\end{figure*}

\begin{figure*}
    \centering
    \includegraphics[width=1.5\columnwidth]{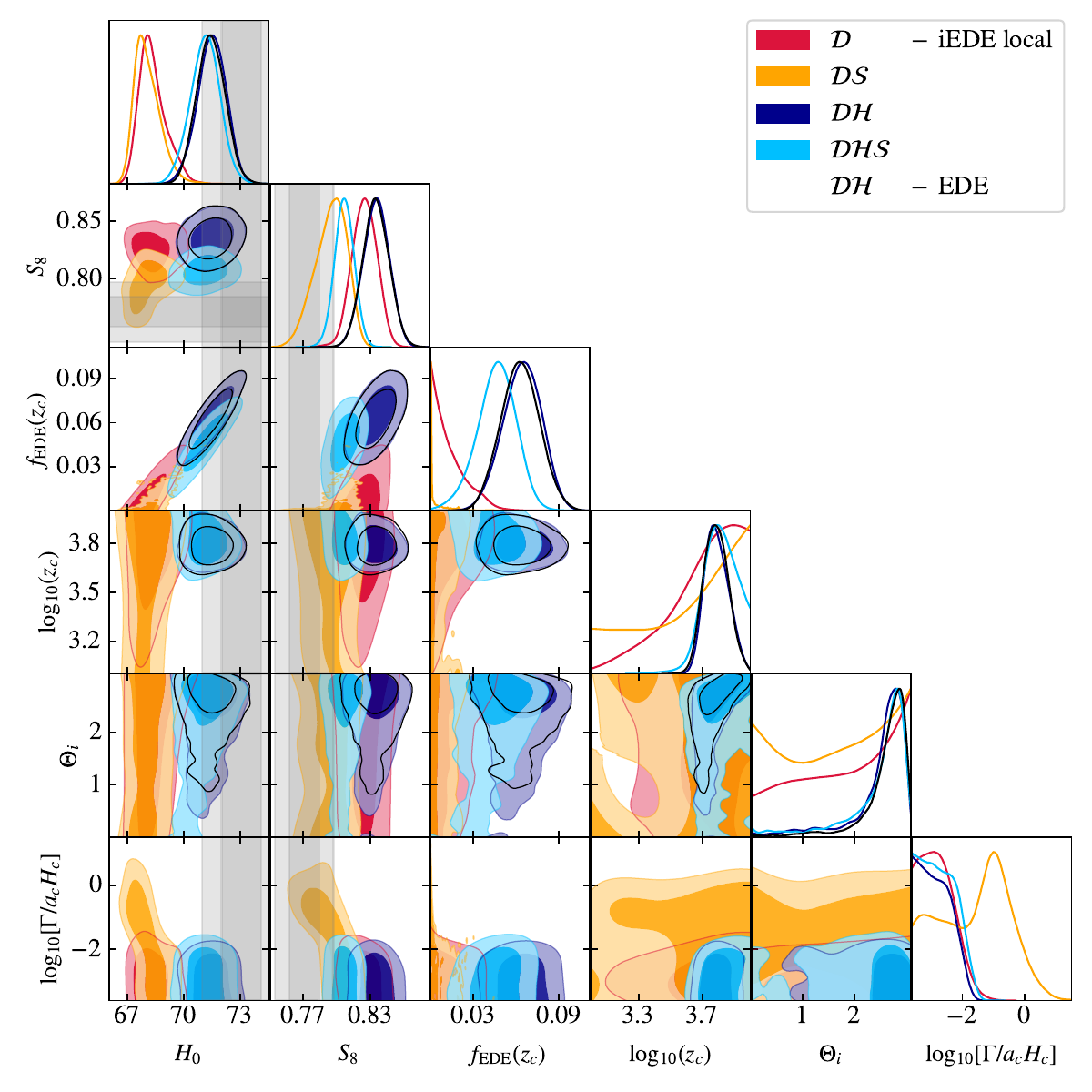}
    \caption{2D posterior distributions reconstructed from the $\mathcal{D}$, $\mathcal{DH}$, $\mathcal{DS}$ and $\mathcal{DHS}$ datasets for the local-interaction EDE scenario. For comparison, we also display the standard EDE 2D posterior distributions reconstructed from the $\mathcal{DH}$ dataset. The gray bands show the $H_0$ and $S_8$ priors defined in Sec.~\ref{sec:analysis}.}
    \label{fig:iEDE_local}
\end{figure*}

In this appendix, we provide all the necessary material for the $i$EDE early and $i$EDE local scenarios: the cosmological constraints are displayed in Tabs.~\ref{tab:cosmoparam} and ~\ref{tab:cosmoparams_early_local}, while the 2D posterior distributions reconstructed from the $\mathcal{D}$, $\mathcal{DH}$, $\mathcal{DS}$ and $\mathcal{DHS}$ datasets are displayed in Figs.~\ref{fig:iEDE_early} and~\ref{fig:iEDE_local}.

\begin{table*}[]
    \centering
    \begin{tabular}{|l|c|c|c|c|}

        \hline
        \multicolumn{5}{|c |}{EDE}\\
        \hline
        \hline\rule{0pt}{3ex}
        & $\mathcal{D}$ &  $\mathcal{DH}$ & $\mathcal{DS}$ & $\mathcal{DHS}$ \\
        \hline

$f_{\rm EDE}(z_c)$
	 & $<0.037(0.016)$
	 & $0.063(0.069)\pm 0.013$
	 & $< 0.022(0.007)$
	 & $0.045(0.050)\pm 0.012$
	 \\
$\log_{10}(z_c)$
	 & unconstrained$(3.96)$
	 & $3.784(3.828)\pm 0.076$
	 & unconstrained$(3.98)$
	 & $3.79(3.88)^{+0.20}_{-0.17}$
	 \\
	 
$\Theta_i$
	 & unconstrained$(3.08)$
	 & $> 1.52(2.87)$
	 & unconstrained$(3.04)$
	 & $> 1.37(2.93)$
	 \\
	 
\hline
$H_0$
	 & $68.34(68.45)^{+0.48}_{-0.86}$
	 & $71.43(71.69)\pm 0.74$
	 & $68.61(68.51)^{+0.36}_{-0.56}$
	 & $71.19(71.29)\pm 0.74$
	 \\	 
$\omega{}_{\rm cdm }$
	 & $0.1215(0.1223)^{+0.0011}_{-0.0025}$
	 & $0.1298(0.1311)\pm 0.0030$
	 & $0.1191(0.1191)^{+0.0008}_{-0.0015}$
	 & $0.1247(0.1257)\pm 0.0024$
	 \\
$10^{2}\omega{}_{b }$
	 & $2.248(2.257)^{+0.016}_{-0.019}$
	 & $2.291(2.295)\pm 0.019$
	 & $2.252(2.253)\pm 0.016$
	 & $2.292(2.298)\pm 0.019$
	 \\
$10^{9}A_{s }$
	 & $2.111(2.105)^{+0.027}_{-0.032}$
	 & $2.142(2.148)^{+0.029}_{-0.033}$
	 & $2.084(2.083)^{+0.024}_{-0.030}$
	 & $2.110(2.108)^{+0.027}_{-0.030}$
	 \\
$n_{s }$
	 & $0.9683(0.9701)^{+0.0042}_{-0.0063}$
	 & $0.9880(0.9907)\pm 0.0060$
	 & $0.9687(0.9687)^{+0.0040}_{-0.0045}$
	 & $0.9849(0.9866)\pm 0.0058$
	 \\
$\tau{}_{\rm reio }$
	 & $0.0566(0.0548)^{+0.0064}_{-0.0074}$
	 & $0.0587(0.0595)^{+0.0068}_{-0.0076}$
	 & $0.0533(0.0529)^{+0.0056}_{-0.0073}$
	 & $0.0560(0.0562)\pm 0.0069$
	 \\
	 \hline
$S_8$
	 & $0.827(0.829)\pm 0.010$
	 & $0.835(0.839)\pm 0.012$
	 & $0.8054(0.807)\pm 0.0078$
	 & $0.8067(0.810)\pm 0.0085$
	 \\	 
$\Omega{}_{m }$
	 & $0.3097(0.3105)\pm 0.0052$
	 & $0.3006(0.3009)\pm 0.0048$
	 & $0.3022(0.3032)\pm 0.0043$
	 & $0.2926(0.2938)\pm 0.0041$
	 \\

\hline
\end{tabular}
\caption{ Mean (best-fit) $\pm 1\sigma$ (or $2\sigma$ for one-sided bounds) of reconstructed parameters of the regular EDE model confronted to various datasets.}
\label{tab:cosmoparam}
\end{table*}

\begin{table*}[]
    \centering
    \begin{tabular}{|l|c|c|c|c|}

        \hline
        \multicolumn{5}{|c |}{iEDE early} \\
        \hline
        \hline\rule{0pt}{3ex}
        & $\mathcal{D}$ &  $\mathcal{DH}$ & $\mathcal{DS}$ &  $\mathcal{DHS}$ \\
        \hline

$f_{\rm EDE}(z_c)$
 & $< 0.033(0.010)$
 & $0.064(0.070)\pm 0.014$
 & $< 0.030(0.001)$
 & $0.050(0.059)\pm 0.012$
	 \\
$\log_{10}(z_c)$
 & unconstrained$(3.755)$
 & $3.765(3.822)^{+0.063}_{-0.082}$
 & $> 3.280(3.985)$
 & $3.783(3.809)^{+0.054}_{-0.073}$
	 \\
	 
$\Theta_i$
 & unconstrained$(0.45)$
 & $> 1.17(2.86)$
 & unconstrained$(0.19)$
 & $> 0.84(2.84)$
	 \\

$\log_{10}[\Gamma / a_c H_c]$
 & $< 0.2(-0.81)$
 & $< -0.6(-2.18)$
 & $-0.73(0.95)^{+1.60}_{-0.64}$
 & $-1.07(-0.60)^{+0.69}_{-0.10}$
	 \\

\hline
$H_0$
 & $68.24(68.07)^{+0.43}_{-0.81}$
 & $71.44(71.82)\pm 0.80$
 & $68.46(67.89)^{+0.46}_{-0.80}$
 & $71.19(71.67)\pm 0.71$
	 \\	 
$\omega{}_{\rm idm }$
 & $0.1213(0.1216)^{+0.0012}_{-0.0025}$
 & $0.1303(0.1311)\pm 0.0031$
 & $0.1202(0.1194)^{+0.0011}_{-0.0022}$
 & $0.1268(0.1289)^{+0.0026}_{-0.0030}$
	 \\
$10^{2}\omega{}_{b }$
 & $2.248(2.254)^{+0.017}_{-0.020}$
 & $2.295(2.298)\pm 0.020$
 & $2.258(2.251)^{+0.018}_{-0.022}$
 & $2.304(2.311)\pm 0.018$
	 \\
$10^{9}A_{s }$
 & $2.112(2.114)^{+0.028}_{-0.034}$
 & $2.148(2.145)\pm 0.032$
 & $2.099(2.101)^{+0.027}_{-0.034}$
 & $2.130(2.144)^{+0.028}_{-0.034}$
	 \\
$n_{s }$
 & $0.9685(0.9691)^{+0.0043}_{-0.0063}$
 & $0.9886(0.9915)\pm 0.0063$
 & $0.9714(0.9691)^{+0.0048}_{-0.0064}$
 & $0.9898(0.9969)\pm 0.0063$
	 \\
$\tau{}_{\rm reio }$
 & $0.0567(0.0563)^{+0.0064}_{-0.0076}$
 & $0.0596(0.0597)^{+0.0065}_{-0.0080}$
 & $0.0544(0.0548)^{+0.0061}_{-0.0076}$
 & $0.0576(0.0591)^{+0.0065}_{-0.0075}$
	 \\
	 \hline
$S_8$
 & $0.823(0.817)^{+0.012}_{-0.011}$
 & $0.832(0.836)^{+0.013}_{-0.011}$
 & $0.795(0.784)^{+0.012}_{-0.010}$
 & $0.8001(0.7976)\pm 0.0099$
	 \\	 
$\Omega{}_{m }$
 & $0.3102(0.3125)\pm 0.0054$
 & $0.3016(0.3000)\pm 0.0051$
 & $0.3062(0.3092)\pm 0.0056$
 & $0.2970(0.2971)^{+0.0042}_{-0.0050}$
	 \\

\hline
\end{tabular}

\vspace{0.5cm}

    \begin{tabular}{|l|c|c|c|c|}

        \hline
        \multicolumn{5}{|c |}{iEDE local} \\
        \hline
        \hline\rule{0pt}{3ex}
        & $\mathcal{D}$ &  $\mathcal{DH}$ & $\mathcal{DS}$ &  $\mathcal{DHS}$ \\
        \hline

$f_{\rm EDE}(z_c)$
	 & $< 0.037(0.021)$
	 & $0.064(0.07)^{+0.014}_{-0.013}$
	 & $< 0.020(2\times 10^{-4})$
	 & $0.046(0.050)^{+0.015}_{-0.012}$ 
	 \\
$\log_{10}(z_c)$
	 & $> 3.26(3.93)$
	 & $3.789(3.846)^{+0.067}_{-0.084}$
	 & unconstrained$(3.43)$
	 & $> 3.65(3.86)$ 
	 \\
	 
$\Theta_i$
	 & unconstrained$(3.09)$
	 & $> 1.06(2.90)$
	 & unconstrained$(1.14)$
	 & unconstrained$(2.91)$
	 \\

$\log_{10}[\Gamma / a_c H_c]$
	 & $< -2.38(-4.24)$
	 & $< -2.63(-3.79)$
	 & $< -0.30(-0.72)$
	 & $< -2.47(-3.91)$  
	 \\

\hline
$H_0$
	 & $68.34(68.81)^{+0.49}_{-0.85}$
	 & $71.48(71.63)^{+0.80}_{-0.70}$
	 & $68.07(67.60)^{+0.48}_{-0.79}$
	 & $71.13(71.50)\pm 0.78$ 
	 \\	 
$\omega{}_{\rm idm }$
	 & $0.1216(0.1231)^{+0.0013}_{-0.0027}$
	 & $0.1302(0.1311)\pm 0.0030$
	 & $0.1194(0.1193)^{+0.0008}_{-0.0011}$
	 & $0.1251(0.1263)\pm 0.0027$ 
	 \\
$10^{2}\omega{}_{b }$
	 & $2.248(2.257)\pm 0.018$
	 & $2.294(2.296)^{+0.020}_{-0.018}$
	 & $2.241(2.232)^{+0.016}_{-0.019}$
	 & $2.295(2.299)\pm 0.020$ 
	 \\
$10^{9}A_{s }$
	 & $2.109(2.114)\pm 0.031$
	 & $2.147(2.140)^{+0.028}_{-0.033}$
	 & $2.097(2.108)^{+0.028}_{-0.033}$
	 & $2.111(2.107)^{+0.027}_{-0.030}$  
  \\
$n_{s }$
	 & $0.9688(0.9731)^{+0.0044}_{-0.0065}$
  & $0.9886(0.9905)^{+0.0065}_{-0.0057}$
  & $0.9668(0.9645)^{+0.0036}_{-0.0052}$
  & $0.9853(0.9883)^{+0.0066}_{-0.0059}$
	 \\
$\tau{}_{\rm reio }$
	 & $0.0562(0.0561)\pm 0.0069$
	 & $0.0602(0.0584)^{+0.0063}_{-0.0079}$
  & $0.0555(0.0577)^{+0.0067}_{-0.0080}$
  & $0.0561(0.0551)\pm 0.0068$ 
	 \\
	 \hline
$S_8$
	 & $0.825(0.831)\pm 0.011$
	 & $0.835(0.838)\pm 0.011$
	 & $0.795(0.785)^{+0.016}_{-0.012}$
	 & $0.8067(0.810)\pm 0.0086$ 
	 \\	 
$\Omega{}_{m }$
	 & $0.3099(0.3090)\pm 0.0053$
	 & $0.3009(0.3015)\pm 0.0045$
	 & $0.3076(0.3114)\pm 0.0063$
	 & $0.2939(0.2932)\pm 0.0044$ 
	 \\

\hline
\end{tabular}
\caption{ Mean (best-fit) $\pm 1\sigma$ (or $2\sigma$ for one-sided bounds) of reconstructed parameters of the $i$EDE early and $i$EDE local models confronted to various datasets.}
\label{tab:cosmoparams_early_local}
\end{table*}

\section{$\chi^2$ tables} 
\label{app:chi2}

In this appendix, we report the best-fit $\chi^2$ per experiment for the $\Lambda$CDM model, the EDE model, as well as the three $i$EDE scenarios for several combinations of data.

\begin{table*}[]
    \centering
    \begin{tabular}{|c|c|c |c c c c c c c c|}
        \hline
        Data & Model & $\chi^2$ tot & P18TTTEE & P18lens & ext-BAO & BOSS & eBOSS & Pan+ & $M_b$ & $S_8$ \\
        \hline \rule{0pt}{3ex}
        \multirow{5}{*}{$\mathcal{D}$} & $\Lambda$CDM & 4405.00 & 2762.10 & 8.86 & 1.22 &  160.27 & 61.20 & 1411.35  & -- & --\\
        &EDE & 4401.85 & 2759.08  & 9.06 & 1.22 &  160.72 & 60.37 & 1411.39 & -- & --\\
        &iEDE early & 4401.70 & 2759.18 & 9.02 & 1.30 & 160.29 & 60.36 & 1411.56 & -- & --\\
        &iEDE local & 4401.17 & 2758.28 & 9.14 & 1.28 & 160.60 & 60.32 & 1411.55 & -- & --\\
        &iEDE late & 4401.30 & 2758.75 & 9.06 & 1.29 & 160.31 & 60.35 & 1411.54 & -- & --\\
        \hline \rule{0pt}{3ex}
        \multirow{5}{*}{$\mathcal{DH}$} & $\Lambda$CDM & 4442.48 &   2766.43  & 9.47 & 2.06 & 158.03 & 60.50 & 1413.34 & 32.65 & -- \\
        &EDE & 4412.38 & 2762.30 & 9.84 & 1.85 & 160.42  & 60.54 & 1413.09 & 4.35 & --\\
        &iEDE early & 4412.34 &  2762.59 & 9.87 & 1.93 & 160.30 & 60.57 & 1413.27 & 3.81 & --\\
        &iEDE local & 4412.34 & 2762.15 & 9.87 & 1.88 & 160.33 & 60.51 & 1413.15 & 4.45 & --\\
        &iEDE late & 4412.16 &  2763.02 & 9.93 & 1.79 & 159.91 & 60.71 & 1412.93 & 3.82 & --\\
        \hline \rule{0pt}{3ex}
        \multirow{5}{*}{$\mathcal{DS}$} & $\Lambda$CDM & 4414.79 &   2764.32 & 10.38 & 1.85 & 158.22 & 60.74 & 1412.85 & -- & 6.43 \\
        &EDE & 4414.14 & 2762.47 & 10.43 & 1.74 & 158.45 & 60.71 & 1412.62 & -- &  7.71\\
       &iEDE early & 4409.89 & 2763.59 & 9.57 & 1.27 & 161.87 & 61.18 & 1411.47 & -- &  0.95\\
        &iEDE local & 4409.95 & 2763.89 & 9.47 & 1.19 &  161.06 & 61.93 & 1411.26 & -- &  1.15\\
        &iEDE late & 4404.18 & 2759.20 & 9.29 & 1.35 & 161.30 & 60.83 & 1411.73 & -- &  0.48\\
        \hline \rule{0pt}{3ex}
        \multirow{5}{*}{$\mathcal{DHS}$} & $\Lambda$CDM & 4447.50 &  2768.46 & 10.86 & 2.52 &  158.16 & 60.77 & 1414.39 & 28.96 & 3.38\\
        &EDE & 4427.87 & 2763.99 & 11.27 & 2.59 & 158.82 & 60.53 & 1414.76 & 6.91 & 8.99\\
        &iEDE early & 4423.10 &  2768.14  & 10.68 & 2.24 & 158.31 & 61.05 & 1413.95 & 4.63 & 4.11\\
        &iEDE local & 4427.73 &  2764.54  & 11.54 & 2.71 & 159.06 &  60.58 & 1415.02 & 5.36 & 8.91\\
        &iEDE late & 4413.59 &  2764.26  & 9.85 & 1.80 & 159.98 & 60.82 & 1412.96 & 3.60 & 0.32\\
        \hline
    \end{tabular}
    \caption{Best-fit $\chi^2$ of the different models considered in this work ($\Lambda$CDM, EDE and $i$EDE) for various combinations of likelihood.}
    \label{tab:chi2}
\end{table*}

\bibliography{biblio}

\end{document}